\documentclass[acmsmall,authorversion,nonacm]{acmart}

\newcommand{\Rule}[1]{\textsc{#1}}

\usepackage{xcolor}      
\usepackage{graphicx}
\usepackage{hhline}
\usepackage{soul}

\usepackage{listings}

\newcommand{\e}[1]{\mbox{\lstinline[basicstyle=\normalsize]|#1|}}
\newcommand{\es}[1]{\mbox{\lstinline[basicstyle=\normalsize]|#1|}}

\usepackage{hyperref}

\usepackage{eiffel}
\usepackage{colortbl}
\usepackage{diagbox}
\usepackage{subcaption}
\usepackage{caption}
\usepackage{enumitem}
\usepackage{bibnames}

\usepackage{rotating}
\newcommand*\rot{\rotatebox[origin=c]{270}}

\renewcommand{\arraystretch}{1.4}

\usepackage{multirow}
\usepackage{float}
\begin{document}
\title{The concept of class invariant in object-oriented programming}


\author{Bertrand Meyer}%
\orcid{https://orcid.org/0000-0002-5985-7434}%
\affiliation{%
	\institution{Constructor Institute}%
	\city{Schaffhausen}%
	\country{Switzerland}%
}%
\affiliation{%
	\institution{Eiffel Software}%
	\city{Santa Barbara}%
	\country{USA}%
}%
\email{Bertrand.Meyer@inf.ethz.ch}

\author{Alisa Arkadova}%
\affiliation{%
	\institution{Previously at University of Toulouse}%
	\city{Toulouse}%
	\country{France}%
}%
\email{allisonark19@gmail.com}

\author{Alexander Kogtenkov}%
\orcid{https://orcid.org/0000-0003-4873-8306}%
\affiliation{%
	\institution{Previously at Constructor Institute}%
	\city{Schaffhausen}%
	\country{Switzerland}%
}%
\affiliation{%
	\institution{Previously at Eiffel Software}%
	\city{Santa Barbara}%
	\country{USA}%
}%
\email{kwaxer@mail.ru}


\begin{abstract}
Class invariants --- consistency constraints preserved by every operation on  objects of a given type --- are fundamental to building, understanding and verifying object-oriented programs. For verification, however, they raise difficulties, which have not yet received a generally accepted solution.
The present work introduces a proof rule meant to address these issues and allow verification tools to benefit from invariants.

It clarifies the notion of invariant and  identifies the three associated problems: callbacks, furtive access and reference leak. As an example, the 2016 Ethereum DAO bug, in which \$50 million were stolen, resulted from a callback invalidating an invariant.

The discussion starts with a simplified model of computation and an associated proof rule, demonstrating its soundness.
It then removes one by one the three simplifying assumptions, each  removal raising one of the three issues, and leading to a corresponding adaptation to the proof rule. The final version of the rule can tackle tricky examples, including ``challenge problems'' listed in the literature.
\end{abstract}



\maketitle



\section{Class Invariants} \label{class_invariants}
The concept of invariant, a property preserved by various operations, plays an important role in many areas of science, particularly mathematics and physics.
Computing is no exception.
Invariants arise on both sides of the basic computing duality: algorithms and data.
For algorithms, reasoning on loops relies on the notion of loop invariant, a property obtained upon loop initialization then preserved by every subsequent iteration \cite{FloydMeaning1967,hoare_69_axiomatic,furia_loop_2014}.
For data, reasoning on object structures relies on the notion of class invariant, obtained upon object creation
then preserved by every subsequent operation on an object. While the general concept of class invariant is clear, going back to a 1972 article \cite{hoare_proof_1972}, there is as yet no widely accepted semantic specification, as exists for loop invariants. The purpose of this article is to provide such a specification.

\subsection{Basics} \label{basics}

Class invariants exist because the values, or ``fields'', making up objects are not arbitrary but often constrained by consistency properties. 
In an object representing a company, assets equal equity plus liabilities (``accounting equation'' \cite{noauthor_accounting_2021}).
Every operation must preserve this property: if it modifies (say) the equity, it must update at least one of the other two fields to ensure that it holds again.

In object-oriented (OO) programming, which structures programs based on the types of their data, or classes, such a property will be part of the invariant for the corresponding \e{COMPANY} class, part of its contract \cite{oosc_1,oosc_2}.
Each operation of the class must, in addition to satisfying its own contract --- assuming its precondition on entry, ensuring its postcondition on exit --- preserve the invariant, part of the contract of the class, which in effect is ``and''-ed to both its pre- and postconditions.
Verification of OO programs relies on these rules.

This basic idea is straightforward, captured by a Groucho Marx quip in a cult scene of \textit{A Night at the Opera} \cite{brothers1935night}: ``\textit{That's in EVERY contract: it's called a sanity clause}''.
The more respectable software name for the ``sanity clause'' is the concept of class invariant, leading to an \textit{Invariant Hypothesis} (section \ref{invariant_hypothesis}) which makes it possible to reason about programs manipulating data structures. Three phenomena, made possible by the constructs of actual programming languages, can invalidate the Invariant Hypothesis, seeming to justify Chico Marx's final retort in the scene: ``\textit{There ain't no sanity clause}''. They are:

\begin{itemize}
    \item
    \textbf{Callbacks}, which can find an object in a state not satisfying the invariant.
    \item
    \textbf{Furtive access} to objects that are temporarily inconsistent.
    \item
   \textbf{Reference leak}, arising from the presence of references (or pointers) and the associated phenomenon of aliasing, which may render an object inconsistent through operations on \textit{other} objects.
\end{itemize}

\noindent Restrictions on the programming language yield a ``\textit{Simple Model}'' of computation (section \ref{simple_model}), with a simple proof rule guaranteeing the Invariant Hypothesis. These restrictions are not acceptable for realistic programming, but removing them can (section \ref{threats_to_sanity_clause}) raise the risks just cited. To remove these risks, sections \ref{callbacks} to  \ref{reference_leaks} successively drop the restrictions, updating the proof rule to preserve the Invariant Hypothesis. The final rule is suitable for verifying actual, unrestricted OO programs and, we hope, restore the faith in a sanity clause. 

\subsection{Contributions and limitations} \label{contrib}

This work builds on the rich literature on class invariants in the past half-century and particularly the last decade (section \ref{related_work}). Two distinctive features are:

\begin{itemize}
    \item \textbf{Proof rule}. Many published discussions propose a ``methodology'' for handling class invariants. We feel that a more rigorous approach is needed in the form of an inference rule, as used in axiomatic (Hoare-Dijkstra-style) semantics to characterize all programming constructs, as in the use of loop invariants in the inference rule for loops.
    \item \textbf{No programmer annotations}. Most approaches require programmers to add special elements, not affecting the program's behavior,  to help verification. In contrast, the proof rule of this article applies to programs as programmers would normally write them, without any such additions.  
\end{itemize}

\noindent Examples of annotations (in approaches reviewed in section \ref{related_work}) include \textit{wrap/unwrap} instructions to express that at a given point objects satisfy, or not, their invariants;   and ``ownership'' type annotations to express that certain objects (such as a list element) can only be accessed through others (such as a list header). They help the verification tools, but complicate the programmers' task and can turn them away from verification. Transferring as much of the burden as possible from programmers to the tools, the intent of the present work, is part of a general effort to make verification part of a normal programming process.

Other contributions include:

\begin{itemize}
    \item A \textbf{theoretical analysis} of the invariant concept (starting with section \ref{simple_model}). To our knowledge, the literature does not so far include a proper explanation of that concept including its subtleties.

           \item A general \textbf{model of OO computation} enabling formal reasoning (again starting with section \ref{simple_model}).
           
           \item As part of this model, the notion of a ``\textbf{Global Consistency Property}'', an overall invariant property of the state of the entire heap.
           
            \item A \textbf{classification of the  issues} (section \ref{threats_to_sanity_clause}), not always properly identified in the literature.
    
        \item A correctness condition, key to avoiding \textit{wrap/unwrap} annotations: objects must \textbf{satisfy their own invariants} before a qualified call (section \ref{callbacks}).
 
    \item The use of \textbf{information hiding mechanisms} and specifically of selective exports to address furtive access (section \ref{slicing_model}). To our knowledge, no approach so far has made the connection between invariant issues and export specifications. Section \ref{smart} removes any undue burden on programmers. 

    \item A simple solution (section \ref{reference_leaks}) to the tricky issue of \textbf{reference leak}, non-optimal but easy to implement in a prover --- along with a more general solution which is harder to integrate in a modular tool.
    
    \item \textbf{Proofs of soundness} for the techniques described (sections \ref{callbacks} to \ref{reference_leaks}).
    \item On the \textbf{notation} side, a concise new way (section \ref{notation}) of expressing inference rules of axiomatic semantics, making it easy to compare candidate rules.
\end{itemize}

\noindent \textbf{Limitations} are discussed in section \ref{conclusion}; in particular, the rule does not yet cover inheritance, which will be addressed in a follow-up, and the soundness proof has not yet been mechanized.

\subsection{Progression and preview} \label{progression}

The main theoretical result of this work is a proof rule, whose final versions appear in sections \ref{reference_leaks:strong} to \ref{reference_leaks:weak_supply}. The discussion starts from a simple version (reflecting the intuitive notion of invariant), which holds in the Simple Model but does not handle the three obstacles that arise in practice; it then refines it in subsequent sections to overcome these obstacles one after the other. The goal of this step-by-step approach is to let the reader understand the precise justification for all components of the rule and the subtleties of the concepts involved.  

For the reader who would like to have a preview of the overall result, the key inference rule derived from this discussion is (\ref{reference_leaks:weak_supply}):

\begin{figure}[H]
    \centering
    \begin{tabular}{c}
  \e{\{Pre$_r$\ (f)} \ $\land$ \e{INV_LOC / r}  \ $\land$ \e{f.}\e{INV}\e{ / r}\e{\}\ \ \ body$_r\hspace{0.3em}$\ \ \ \{Post$_r$\ (f) } \ $\land$ \e{INV / r}  \ $\land$ \e{f.INV}\e{\}\}} \\    \hline
    \e{\{x.Pre$_r$\ (a)} \ $\land$ \e{INV} \es{/ r}  \ $\land$ \e{a.}\e{INV}\ \e{ / r}\e{\}\ \ x.r (a)}\ \ \e{\{x.Post$_r$\ (a) } \ $\land$ \e{x.INV /r}  \ $\land$ \e{a.INV}\e{\}\}} \\
    \end{tabular}
    \caption*{\Rule{Weak-S}}
    \label{tab:final_invariant_rule}
\end{figure}

\noindent characterizing the semantics of a call \e{x.r (a)} to a routine \e{r} with body \e{body$_r$} and formal arguments  \e{f}. \e{Pre$_r$} and \e{Post$_r$}, applicable to both \e{f} and the call's actual arguments \e{a}, are \e{r}'s pre- and postconditions. \e{INV} denotes the invariant of respective classes (client and supplier); it can be applied to specific objects, as in \e{x.INV}. \e{INV_LOC} is the part of \e{INV} that only depends on an  object's fields, not on other objects.  Finally, \e{INV / r} is the result of ``slicing'' an invariant by a routine: removing all  clauses of \e{INV} that have more visibility, in the sense of information hiding, than \e{r}. (Section \ref{smart} will introduce a small optional optimization allowing slightly more liberal versions of the conditions \e{INV /r } and \e{a.INV / r}.)

This rule has variants and complements (for initialization, for pure queries), but is the basis for proofs of the literature's ``challenge problems'' in section \ref{challenge_problems_and_solutions}.

\subsection{Terminology} \label{terminology}

Presentations of object-oriented programming vary in their terminology, often influenced by a specific programming language. The present article uses the following consistent terminology, going back to \cite{meyer_eiffel_1988}:
\begin{itemize}
\item
An \textit{object} is a run-time instance of a \textit{class}. (``Class'' is a static concept, denoting a part of the program text, and ``object'' a dynamic concept.)
\item
\textit{Type} means the same as ``class'' for this discussion, so that ``an object of type \e{T}'' means the same as ``instance of \e{T}'', and a program may declare a variable ``of type \e{T}'' to specify that its run-time values will be references to instances of \e{T}. (``Type'' is in fact a more general concept, particularly in the presence of generic classes, but the differences do not matter for this article. More generally, typing, static or dynamic, plays no role in the discussion.) 
\item
An object is a collection of individual values or \textit{fields}. In the corresponding class, each field   is defined by an \textit{attribute} (sometimes also called field, but it is better to distinguish the static and dynamic concepts).
\item
Classes define operations, or \textit{features}, on their instances. A feature can be a \textit{command}, also known as a procedure, which can change one or more objects, but does not return a result; or a \textit{query}, which returns information about the object. (In C++, the term for ``feature'' is ``member''.)
\item
A query (such as ``obtain a bank account's balance'') can be implemented as an \textit{attribute} (look up the \e{balance} field in the object) or a \textit{function} (define an algorithm to compute the balance from successive deposits and withdrawals). 
\item
A \textit{routine}, or ``method'', is an algorithm applicable to the instances of a class; routines cover both procedures and functions.
\end{itemize}

\noindent We distinguish between the ``class invariant'', a consistency constraint defined in a class text, applying to all instances of the class, and an ``object invariant'', its application to a given instance. (For no apparent reason, some work cited in section \ref{related_work} uses ``object invariant'' to mean ``class invariant'', even though the latter term has been around for 50 years.) Note that a class invariant, and corresponding object invariants for instances, denote a property of the fields of a \textit{single} object, even though some of these fields can be references to other objects and the invariant may involve queries on them. A property that describes the combined consistency of several objects is called a \textit{multi-object invariant}.

\section{Working with invariants} \label{working_with_invariants}

Class invariants are a tool for reasoning about the correctness of programs manipulating possibly complex run-time data structures. We review the concept, first informally then through an initial proof rule which captures the essential simplicity and elegance of the class invariant concept.

\subsection{Working with heaps}

\begin{figure}[h]
    \centering
    \includegraphics[scale=0.24]{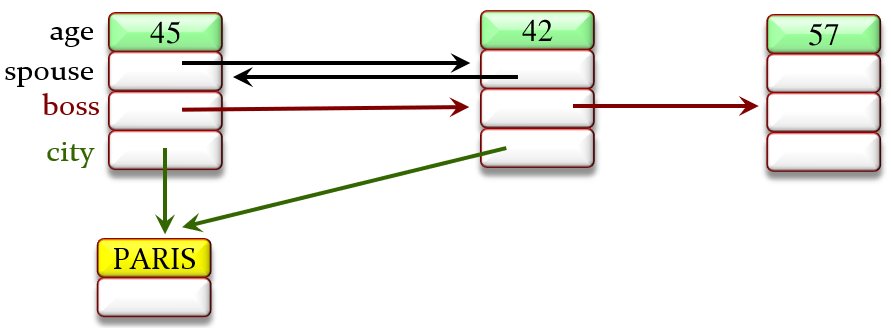}
    \caption{Some objects in a heap}
    \label{fig:heap}
\end{figure}

A run-time object structure is made of objects with mutual references and is called a heap. Fig. \ref{fig:heap} shows an illustration of a small part of a possible heap.

\noindent Objects are, as noted, made of ``fields''. Some fields, such as \e{age}, are basic values such as integers and strings, while others, such as \e{spouse}, are references to other objects. We say that such an object is \textit{attached} to the reference as represented in the program; for example, the bottom object in Fig. \ref{fig:heap} is attached to the \e{city} field of the top-left object. The objects in the figure represent three persons and one city; some fields have been left blank.

\begin{figure}[h]
\centering
    \includegraphics[width=0.7\textwidth]{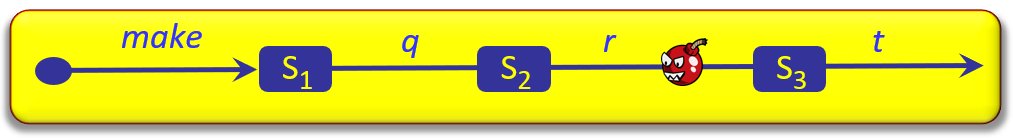}
    \caption{Object lifecycle}
    \label{fig:object_lifecycle}
\end{figure}

A ``qualified call'' \e{x.r (a)} executes a routine \e{r} on a target object denoted by \e{x}, with (optional) arguments \e{a}. We may view the life of a single object S during execution as a sequence of such calls, illustrated by Fig. \ref{fig:object_lifecycle}. After an initial creation, through a ``constructor'' or ``creation procedure'' called \e{make} in the figure, the object executes successive operations (\e{q, r, t}) as a result of qualified calls from client objects (which will appear in a more complete version of the figure, Fig. \ref{fig:oo_model} in section  \ref{dynamic_model}). Fig. \ref{fig:object_lifecycle} shows the successive states (S$_1$, S$_2$, S$_3$, ...) of the object: after creation (\e{make}), then before and after every qualified call (to routines \e{q}, \e{r}, \e{t}, ...).

\noindent Every object is (section \ref{terminology}) an instance of a class defined in the program text. The objects in Fig. \ref{fig:heap} are assumed to be instances of classes \e{PERSON} and \e{CITY}. Such classes typically specify consistency constraints on their instances, which make up the class invariant. For example (section \ref{basics}), the invariant of a \e{COMPANY} class may specify \e{assets = equity + liabilities}. In Fig. \ref{fig:heap}, the invariant of class \e{PERSON} may include \e{city = spouse.city}  (I live in the same city as my spouse) and \e{spouse.spouse = Current} (the spouse of my spouse is myself, with \e{Current}  denoting the current object as explained below).

\subsection{The Invariant Hypothesis and the Scandalous Obligation} \label{invariant_hypothesis}

Invariants such as the ones cited are crucial for reasoning about programs. In spite of the name, a class invariant does not need to hold at all times; it can be temporarily violated during a routine's execution, as at the marked point in \e{r}'s execution in Fig. \ref{fig:object_lifecycle}. But it must hold in states (S$_1$ etc.) where client objects can start a new call. An object that satisfies its invariant is said to be \textbf{consistent}.

This fundamental property is the ``Invariant Hypothesis'':   

\begin{center}
    \begin{tabular}{|c|}
     \hline
     \textcolor{blue}{\textbf{Invariant Hypothesis (initial)}}  \\
     At the start of every qualified call, the target object is consistent.\\
     \hline
     \end{tabular}
\end{center}

\noindent (``Initial'' because a slightly revised version will appear in \ref{relative_consistency}.) The ``target'' in \e{x.r (a)} is the object denoted by \e{x}. A ``consistent'' object satisfies the invariant of its class. The Invariant Hypothesis enables both program construction and verification to rely on the class invariant, by assuming that every object satisfies it whenever it is accessible from the rest of the system (in states S$_1$, S$_2$ ...). We deduce this property simply by showing that every creation procedure (constructor) of the class ensures it (in state S$_1$ in Fig. \ref{fig:object_lifecycle}) and every exported feature (every feature \e{r}  that clients can use, in calls of the form  \e{x.r (a)}) preserves it.

This mode of reasoning takes after mathematical induction (of which invariants, whether for classes or loops, are the application to computing).
To deduce a property $P$ of all natural numbers it suffices to prove that $P (0)$ holds and that $P (n + 1)$ follows from $P (n)$ for any $n$.
Then we can assert, for example, $P (386)$.
We do not need to go back to $P (385)$, then $P (384)$ and so on.

To understand the importance of the Hypothesis, one may consider what happens if it does not hold. The invariant would just be a clause repeated explicitly in the pre- and postconditions of every routine, which would be expressed as \e{Pre $\ \land\ $ INV} and \e{Post $\ \land\ $ INV}, where \e{INV} captures common elements. But then every caller would have to establish \e{x.INV}  before every call \e{x.r (a)}. We call this transfer of responsibility from a class to its clients the \textbf{Scandalous Obligation}.

The Scandalous Obligation would render the notion of invariant largely useless (it is the equivalent of forcing the proof of $P (386)$ to go over all previous values). Also note that the invariant may include secret (private) features, which clients cannot directly affect. The rest of this article shows how to ensure the Invariant Hypothesis, which unleashes the induction-like power of the invariant concept by enabling clients always to assume, prior to a call, that the supplier object is consistent (satisfies its invariant). A model for OO computation, such as those appearing in sections \ref{simple_model} to \ref{reference_leaks}, will only be considered \textbf{sound} if it guarantees the Hypothesis. 

An issue that does not arise with mathematical induction is that the Invariant Hypothesis only talks about the \textit{start} of qualified calls: what about their \textit{end}? As Fig. \ref{fig:object_lifecycle} suggests, nothing happens to an object between the end of a qualified call of  target S and the start of the next call on S; but something could happen to \textit{other} objects that invalidates S's invariant (\textit{reference leak}, studied in section \ref{reference_leaks}). Separately from the Invariant Hypothesis, the postconditions of all rules will ensure that objects are consistent on call completion too.

We now turn to a formal expression of these concepts.

\subsection{Background: a classical proof rule for routine calls} \label{classical_context}

Axiomatic reasoning on programs uses ``Hoare-style'' inference rules such as this one for non-OO routine calls (ignoring recursion and termination)  \cite{engeler_procedures_1971}:
\begin{figure}[H]
    \centering
    \begin{tabular}{r c l}
    \e{\{Pre$_r$(f)\}} & \e{body$_r$} & \e{\{Post$_r$(f)\}} \\
    \hline
    \e{\{Pre$_r$ (a)\}} & \e{r (a)} & \e{\{Post$_r$ (a)\}}
    \end{tabular}
    \Description{Classical inference rule for an unqualified routine call involving only preconditions and postconditions}
    \caption*{\Rule{Classic}}
    \label{tab:classic_rule}
\end{figure}
\noindent Such a rule is a proof technique: below the line appears a property of a construct, here  the call \e{r (a)}; the rule states that we may deduce it by proving the property above the line, applicable to the body of \e{r} (\e{f} denoting its formal arguments). The properties involved are ``Hoare triples'' of the form \e{\{P\} A \{Q\}}, meaning that executing \e{A} in a state satisfying \e{P} will yield one satisfying \e{Q}.

\Rule{Classic} states that the effect of a call is that of executing the body after substituting actual for formal arguments.
It captures the role of routines (methods): abstracting a computation by giving it a name and parameterizing it.

In an OO context, \Rule{Classic} remains applicable to unqualified calls \e{r (a)}, executed on the current object (see \ref{qualified}).
Also, if we ignore the invariant, to qualified calls:
\begin{figure}[H]
    \centering
    \begin{tabular}{r c l}
    \e{\{Pre$_r$(f)\}} & \e{body$_r$} & \e{\{Post$_r$(f)\}} \\
    \hline
    \e{\{x.Pre$_r$ (a)\}} & \e{x.r (a)} & \e{\{x.Post$_r$ (a)\}}
    \end{tabular}
    \Description{A rule for a qualified routine call that does not take class invariant into account}
    \caption*{\Rule{No-Invariant}}
    \label{tab:no_invariant_rule}
\end{figure}
\noindent The \Rule{No-Invariant} rule is conceptually the same as \Rule{Classic}, but takes into account the role, in OO programming, of the target of a qualified call, here \e{x}, which it simply treats as if it were an extra argument.

\subsection{Notation} \label{notation}

The rest of the discussion will introduce variants of the rule \Rule{No-Invariant}. Since they have many elements in common, we can rely on a concise tabular notation to avoid repetitions and emphasize the distinctive novelty of each variant. This notation expresses \Rule{No-Invariant} as the highlighted part of the following table (the rest provides explanations, not repeated in later rules):
\begin{figure}[H]
    \centering
    \setlength\extrarowheight{-2pt}
    \begin{tabular}{|l|l|l|}
    \cline{1-3}
    \multirow{2}{*}{Notes} & \multicolumn{1}{l|}{Precondition} & Postcondition  \\
     & \multicolumn{1}{l|}{(\es{P} in \es{\{P\} A \{Q\}})} & (\es{Q} in \es{\{P\} A \{Q\}}) \\  \hhline{-|-|-}
     
    Proof obligation for & \cellcolor{green!25} & \cellcolor{green!25}  \\
    exported routine \es{r} with  & \cellcolor{green!25}\es{Pre (f)} & \cellcolor{green!25}\es{Post (f)}  \\
   formal arguments \es{f} in  & \cellcolor{green!25} & \cellcolor{green!25}  \\
   supplier class \es{SC} & \cellcolor{green!25} & \cellcolor{green!25}  \\ \hline \hline
    
    Call rule for a call & \cellcolor{green!25}& \cellcolor{green!25} \\
    \es{x.r (a)} in client class & \cellcolor{green!25}\es{x.Pre (a)} & \cellcolor{green!25}\es{x.Post (a)}  \\
    \es{CC}, with \es{x} of type \es{SC} & \cellcolor{green!25} & \cellcolor{green!25} \\ \cline{1-3}
    \end{tabular}
    \caption*{\Rule{No-Invariant} rule in tabular notation (highlighted part only)}
    \label{tab:notation}
\end{figure}
\noindent Since the context is always the same, the notation further drops \e{r} (i.e. \e{Pre} stands for \e{Pre$_r$} and so on).
On the supplier side, \e{INV} will stand for the invariant of class \e{SC};
on  the client side, for the invariant of class \e{CC}. Since \e{x} is of type \e{SC}, \e{x.INV} stands for the invariant of \e{SC} applied to \e{x}.

\subsection{Sound but useless invariant integration} \label{sound_useless}
\Rule{No-Invariant} does not recognize the role of the invariant; the author of the code would have to add it manually to \e{Pre} and \e{Post} for every exported routine \e{r}).
Rule \Rule{Pointless} is a first attempt at integrating the invariant explicitly:
\begin{figure}[H]
\centering
        \begin{tabular}{|ll|ll|}
        \cline{1-4}
        BS$_1$: & \es{INV}        & AS$_1$:    & \es{INV} \\
       BS$_2$:    & \es{Pre (f)}    & AS$_2$:      & \es{Post (f)}          \\ \hline \hline
        BC$_1$: & \es{x.INV}      & AC$_1$:    & \es{x.INV}                 \\
        BC$_2$:   & \es{x.Pre (a)}  & AC$_2$:      & \es{x.Post (a)}            \\ \cline{1-4}
        \end{tabular}
    \caption*{\Rule{Pointless}}
    \label{tab:pointless_rule}
\end{figure}
\noindent (Multiple assertions in a box, such as \e{INV} and \e{Pre} in BS$_1$ and BS$_2$, are to be combined by conjunction.
In mnemonics for entries, BS$_1$ etc., B stands for before and A for after, S for supplier and C for client.)

This rule and the ones that follow only talk about preservation of the invariant for an object that already satisfies it as a result of proper initialization. Section \ref{initialization_queries_concurrency:initialization} will introduce a creation rule covering initialization.

\subsection{The Invariant Preservation Property} \label{IPP}

The reason for calling the previous rule \Rule{Pointless} is that it fails to take advantage of the invariant.
It is in fact the same as \Rule{No-Invariant}, with pre- and postconditions extended with the invariant.

The disappointment is clause BC, requiring clients to obtain \e{x.INV} prior to calls.
This approach suffers from the Scandalous Obligation (\ref{invariant_hypothesis}), losing the basic idea of invariants expressed by the Invariant Hypothesis. To satisfy the Hypothesis, we need a rule in which the invariant for the target object:
\begin{itemize}
    \item Will hold after the call (as represented by the presence of \e{x.INV} in AC).
    \item May be assumed to hold before the call (BC), without transferring the responsibility to the caller.
\end{itemize}

\noindent Under such assumptions, one may prove the correctness of any call, in the form

\begin{lstlisting}[xleftmargin=3em]
{x.Pre (a)} x.r (a) {x.Post (a) $\land$ x.INV}
                -- Call Correctness Property (CCP)
\end{lstlisting}

\noindent with no \e{x.INV} on the precondition side, simply by having proved once and for all, independently of any callers, the correctness of the routine, in the form

\begin{center}
    \begin{lstlisting}[xleftmargin=3em]
{Pre (f) $\land$ INV} body$_r$ {Post (f) $\land$ INV}
                -- Invariant Preservation Property (IPP)
\end{lstlisting}
\end{center}

\noindent \textbf{Proving} \textbf{a class correct} means proving the IPP (plus initialization, see \ref{initialization_queries_concurrency:initialization}) for its exported routines, which then suffices (this is what the Invariant Hypothesis means) for proving the correctness of any call (CCP, \textit{without} \e{x.INV}  on the left).


\subsection{The ideal rule} \label{ideal_rule}

Removing \e{x.INV} in BC gives an ``ideal'' rule, which satisfies the Invariant Hypothesis and in a simpler world would be the final word:
\begin{figure}[H]
\centering
    \begin{tabular}{|cc|cc|}
    \cline{1-4}
    BS: & \e{INV}        & AS:    & \e{INV}     \\ \hline \hline
    BC: & \cellcolor{green!25}     & AC:    & \e{x.INV}    \\ \cline{1-4}
    \end{tabular}
    \caption*{\Rule{Ideal}}
    \label{tab:ideal_rule}
\end{figure}

\noindent (Since the \e{Pre} and \e{Post} parts are not specific to OO programming and apply equally to all rule variants, we omit
them from now on.) The novelty is the highlighted empty BC entry, which does not contain the requirement \e{x.INV} on the caller. If the \Rule{Ideal} rule applies, the callers to a routine satisfying the CCP:
\begin{itemize}
    \item \textit{Never} have to worry about guaranteeing \e{x.INV} on \textit{entry} (as this property was ensured on creation, then preserved by previous operations); that is, are spared the Scandalous Obligation.
    \item Are \textit{always} able to rely on \e{x.INV} on \textit{exit} (since re-establishing this sanity clause is part of the contract of \e{r}).
\end{itemize}

\noindent The \Rule{Ideal} rule is the direct formal expression of the Invariant Hypothesis, and the formal embodiment of the notion of class invariant. (The antecedent of the rule --- BS and AS --- expresses the Invariant Preservation Property.) A ``Simple Model'' of computation, described in the following section, satisfies it. The vagaries of actual programming in realistic languages will threaten to bring back the Scandalous Obligation and require adjustments to the rule.

\subsection{The run-time model} \label{dynamic_model}

\begin{figure}[h]
    \centering
    \includegraphics[scale=0.2]{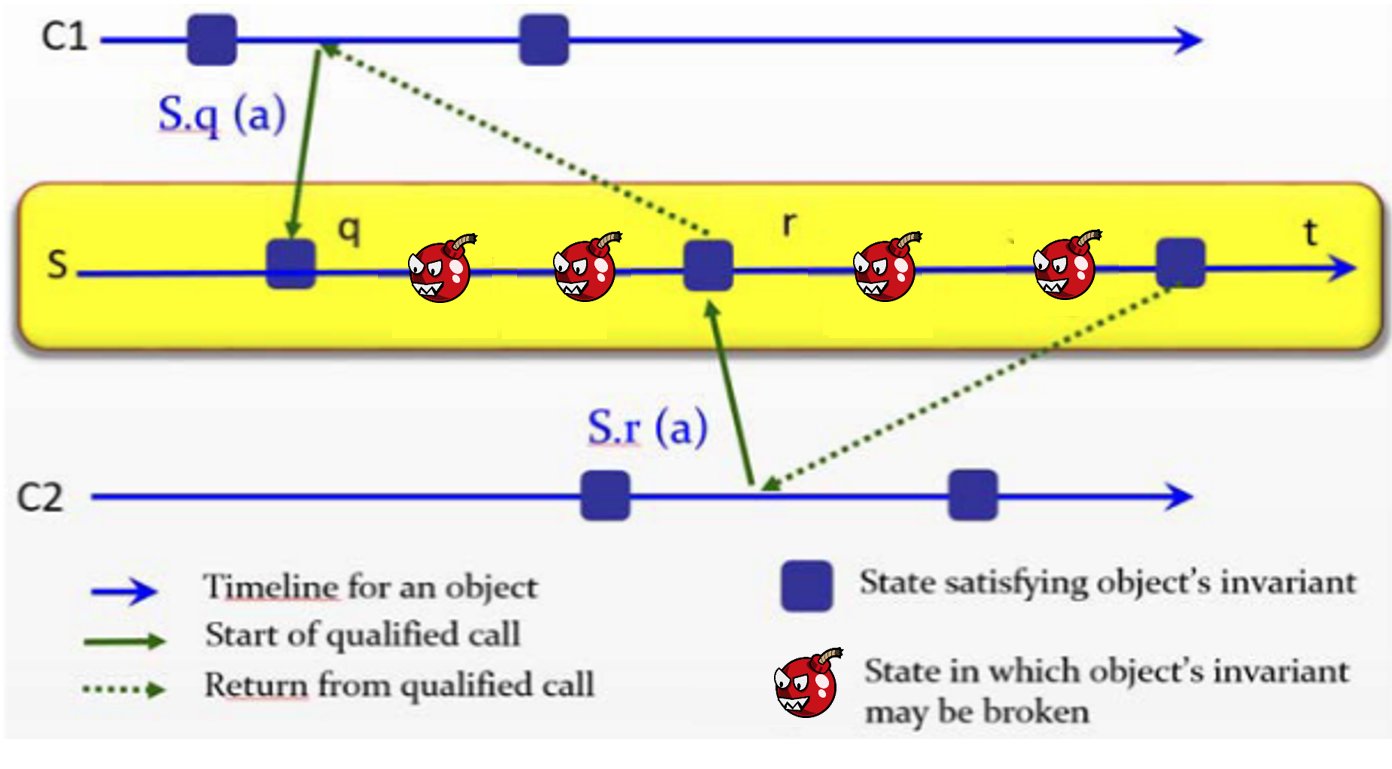}
    \caption{OO computation model}
    \label{fig:oo_model}
\end{figure}

To understand how the \Rule{Ideal} rule applies and where it requires adaptation, one must put the lifecycle of a single object (Fig. \ref{fig:object_lifecycle}) in the broader context of an entire OO system's execution, involving any number of objects (as in the heap of Fig. \ref{fig:heap}). Fig. \ref{fig:oo_model} includes, along with the original supplier object S, the history --- after initialization --- of two of its client objects, C1 and C2. 

\noindent As in the earlier figure, \e{S} executes routines \e{q}, \e{r}\ldots; Fig. \ref{fig:oo_model} shows that these executions result from qualified calls coming from clients \e{C1} and \e{C2}.
\e{C1} calls \e{q} on \e{S}, and at some later time \e{C2} calls \e{r} on \e{S}.

\subsection {Current object, active objects, active routines}
\label{current_active}

At any point during execution, one of the objects is the ``current object'', known as \e{Current} (or ``this'' in languages such as Java).
Each execution step is either:
\begin{itemize}
    \item An internal operation, which only affects the state of the current object.
    An example is an assignment modifying one of the object's fields.
    \item An external operation, in the form of a qualified call \e{x.r (a)}, which triggers a routine on another object, denoted by \e{x}.
    That object becomes current for the duration of the routine's execution.
\end{itemize}

\begin{figure}[h]
    \centering
    \includegraphics[scale=0.4]{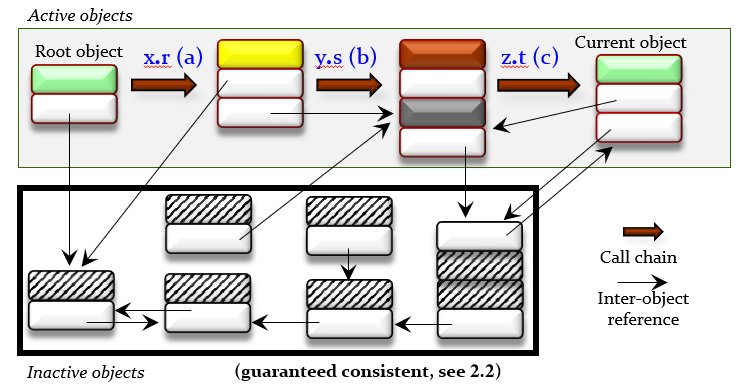}
    \caption{Run-time snapshot: root, current, active and inactive objects}
    \label{fig:snapshot_simple}
\end{figure}

\noindent Execution starts with a ``root'' object, the first ``current''.
As a result of qualified calls, other objects become current. At any time, objects are partitioned into two categories (Fig. \ref{fig:snapshot_simple}): objects in the call chain, starting from the root down to ending with the current object, called \textbf{active}; all others, called inactive. (In concurrent programming, execution may include more than one root and more than one call chain.) Note that an object \e{S} is active if it has started and not yet completed a qualified call \e{x.r (a)}. We call \e{r} the ``routine of'' the active object; \e{r} and other routines along the call chain are the ``active routines''. Note that the current object is active, although the semantic rules will treat it differently from other active objects. Inactive objects appear in the lower part of Fig. \ref{fig:snapshot_simple}.

As suggested in that figure, fields of objects in any category may be references to objects of any other.

\subsection {Consistent and inconsistent objects} \label{invariant_satisfaction}

We saw in \ref{invariant_hypothesis} that during execution some objects may temporarily become inconsistent (violate their invariants).  The objects in Fig. \ref{fig:oo_model} go from consistent state to consistent state (blue squares), traversing inconsistent states in-between. In particular, internal operations often break the invariant: with the invariant clause \e{assets = equity + liabilities}, after a change to \e{equity} the invariant no longer holds. The routine must restore it by changing one or both other fields.

The key formal property of an OO computational model is the specification, represented by a thick black rectangle in Fig. \ref{fig:snapshot_simple}, of which objects \textit{must} be consistent. It will be known as a  \textbf{Global Consistency Property} or GCP. Fig. \ref{fig:snapshot_simple} illustrates version GCP-0, whose definition in  \ref{GCP-0} below will state that among non-current objects \textit{only active objects are permitted to be inconsistent}. They include the current object, which in all models can  be inconsistent since it may be executing an internal operation. Later models with their associated proof rules, representing increasingly sophisticated forms of OO computation, use different variants, GCP-1 to GCP-3. For each variant, we will need to prove two results:

\begin{itemize}
\item A \textbf{Global Consistency Lemma} stating that every run-time operation preserves the relevant GCP.

\item A \textbf{Soundness Theorem} (following from the Lemma), stating that the rule's conclusion follows from its antecedent, and that the Invariant Hypothesis holds (every qualified call finds its target in a consistent state).
\end{itemize}

\noindent (A Global Consistency Property is a \textit{global invariant} on the heap. To avoid confusion, the discussion will  reserve ``invariant'' for class and object invariants.)

\subsection{Qualified and unqualified calls} \label{qualified}

Along with qualified calls \e{x.r (a)}, OO languages support \textbf{unqualified} calls \e{r (a)}, which apply a routine \e{r}  to the current object. To avoid any confusion (sometimes found in the literature), note that \textbf{the rules on class invariants apply to qualified calls only}. An unqualified call \e{r (a)} can be understood  (``inlined'') as the sequence of instructions making up the body of r (accounting for possible recursion), and is not subject to the invariant: it may not assume it on entry, and does not have to ensure it on exit. Accordingly, the applicable rule for such a call is the traditional non-OO call rule: \Rule{Classic} from section \ref{classical_context}, applied to the current object. What determines the rule to apply is not whether \e{r} is exported, but whether a given \textit{call} is qualified, \e{x.r (a)}, or unqualified, \e{r (a)}. (Note that \e{Current.r (a)} and \e{r (a)}, which perform the same computation, are bound by different rules: the first is subject to the invariant, the second is not.)

A routine \e{r} involved in an unqualified call can, of course, include a qualified call, but that call is only relevant for the process of proving \e{r} itself correct.

\section{The Simple Model} \label{simple_model}

What determines whether the \Rule{Ideal} rule and Invariant Hypothesis hold is both what properties invariants may state about objects and how operations (in the computational model presented above) may change them. The ``Simple Model'' of this section meets the rule and the hypothesis by severely restricting the expressive power of OO programs. Its proof of soundness in the rest of the section corresponds to proofs about simplified OO models in the literature.

\subsection{The restrictions} \label{restrictions}

The Simple Model makes three assumptions:
\begin{center}
    \begin{tabular}{|l|}
     \hline
     \textcolor{blue}{\textbf{Assumptions in the Simple Model}}  \\
     \hline
        \textbf{[SM1]} There is no callback. (``\textit{No-callback}''.)\\
        \textbf{[SM2]} Invariant Preservation Property (\ref{IPP}): every routine that can be used  in a \\
        \ \ \ \ \ \  qualified call preserves the class invariant (``\textit{Invariant preservation}''.)\\
        \textbf{[SM3]} Every invariant is a function of
                the object's fields only. (``\textit{Invariant locality}.'') \\
     \hline
    \end{tabular}   
\end{center}
\noindent A ``callback'', ``direct'' or ``indirect'', is a qualified call whose target is (respectively) the current object or another active object. (An illustration of an indirect callback will appear in Fig. \ref{fig:callback_to_active}, section  \ref{slicing_condition}.)  Assumption SM1 rules out both kinds.

Assumption SM2, Invariant Preservation, expresses the invariant concept in its simplest form. It is guaranteed by the antecedent part of the \Rule{Ideal} rule (\ref{ideal_rule}), BS and AS (Invariant Preservation Property as seen in \ref{IPP}).

SM3, Invariant Locality, excludes any invariant clause \e{x.some_property} where \e{x} is a reference to another object: the invariant depends on the fields of a single object. A clause such as \e{assets = equity + liabilities}, involving only integer fields, satisfies this condition, but not any of the example invariants on \e{PERSON}, such as \e{spouse.spouse = Current}, which involves another object. These examples indicate that the restriction is severe.

A formal OO framework consists of a programming language (with any associated restrictions), a computational model, and Hoare-style inference rules for qualified calls (and initialization as will be added in section \ref{initialization_queries_concurrency:initialization}). We say that the framework is sound if the computational model satisfies the rules.

We will now prove that the Simple Model with the \Rule{Ideal} rule is sound.

\subsection {Global consistency property} \label{GCP-0}

The Simple Model's key property is that it preserves Global Consistency, version GCP-0 (\ref{invariant_satisfaction}, see Fig. \ref{fig:snapshot_simple}): \textbf{at all times, all inactive objects are consistent}.

\textbf{Global Consistency Lemma}: in the Simple Model, all run-time events preserve GCP-0.

\textit{Proof}: while specifying the full semantics of a program's execution requires a formal model of the entire language, the analysis of invariant-related behavior need only consider run-time events of three kinds (recognizable in Fig. \ref{fig:object_lifecycle}):
\begin{itemize}
\item \textit{Internal operation}: perform an internal operation on the current object.
\item \textit{Start-of-call}: start a qualified call \e{x.r (a)}.
\item \textit{End-of-call}: terminate such a call, returning to the calling object and routine.
\end{itemize}
The following table characterizes each in terms of: (1) whether it affects which object is ``current''; (2) whether it affects the invariant of any object.

\begin{center}
      \renewcommand{\arraystretch}{1.1}
    \begin{tabular}{|c|c|c|}
         \hline
         Event type & Which object & Possibly invalidated \\
                    & becomes current? & invariants \\
         \hline
         \hline
         Internal operation & No change (current & \cellcolor{green!25}Invariant of \\
                            & remains current) & \cellcolor{green!25}current object \\
         \hline
         Start of call \es{x.r (a)} & Object attached to \es{x} & None \\
         \hline
         End of call & Previously current object & None \\
        \hline
    \end{tabular}
\end{center}
\noindent Events of each kind preserve \textbf{GCP-0}:
\begin{itemize}
    \item [G1] \textit{Internal operation}: because of invariant locality (assumption SM3 in the Simple Model), the only invariant that the operation could invalidate is the invariant of the current object (highlighted entry). GCP-0, which does not require the current object to be consistent, is not affected.
    
    \item [G2] \textit{Start-of-call}: the target (the object attached to \e{x}) becomes current. The previous current and all other active objects up the call chain remain active. Since no object becomes newly inactive, GCP-0, which only constrains inactives, continues to hold. Note that the target (the new current) is consistent anyway since it was inactive by the assumption SM1 of no callbacks.
   
    \item [G3] \textit{End-of-call}: two objects are involved, the target of the call, say \e{S}, and the previously current object, \e{C}. \e{S} becomes inactive again (otherwise the call would have been a callback, violating SM1). As we just saw in G2, \e{S} was consistent before the call. Thanks to SM2 (invariant preservation), it is now consistent again. \e{C} becomes current and hence does not have to satisfy its invariant under GCP-0. $\blacksquare$
    
\end{itemize}
This proof required all three assumptions SM1, SM2,  and SM3 of the Simple Model. Subsequent sections will adapt it as we remove them one by one.

In passing, we have proved (discussion of G2 and G3) that the Simple Model satisfies the \textbf{Consistent Target Theorem} (CTT): on entry and exit of a qualified call, the target is consistent. 

\subsection{The Simple Model is sound} \label{simple_model_sound}

\textbf{Simple Model Theorem}: For the Simple Model, the \Rule{Ideal} rule is sound.

\textbf{Proof}: the property to establish is that if a routine has been proved correct (antecedent of the rule, with \e{INV} in both the pre- and postconditions of the routine, entries BS and AS), the postcondition \e{x.INV} (entry AC) will hold upon completion of any qualified call. This property immediately follows from the CTT theorem above, which also implies that the Invariant Hypothesis holds. $\blacksquare$

This result justifies leaving the BC entry blank in the rule: the client does not need to worry about \e{x.INV} before the call, avoiding the Scandalous Obligation.

An important observation applies to proofs of soundness for \Rule{Ideal}  and its followers. All of them have the postconditions \e{INV} for the supplier (entry AS) and correspondingly \e{x.INV} for the client (AC). Unlike \Rule{Pointless}, however, they do not include the precondition \e{x.INV} for the client (entry BC); \Rule{Ideal} itself has nothing there. If we are able to prove the Invariant Hypothesis, we guarantee that \e{x.INV} holds on start-of-call and hence can simply apply \Rule{Pointless} to derive the postcondition AC. As a result, \textbf{to prove such a model sound it suffices to prove that its satisfies the Invariant Hypothesis}. 

\section{Threats to the Invariant Hypothesis} \label{threats_to_sanity_clause}

Three phenomena can invalidate the properties of the Simple Model in the world of real OO programming: callbacks, furtive access and reference leak.
They are of a different nature (although sometimes commingled in the literature) and subject to different remedies.
Each results from renouncing one of the assumptions of the Simple Model.
Callbacks and furtive access are closely related; reference leak is of a different nature, arising from the use of references to objects in OO programming and the resulting possibility of aliasing.

\textbf{Callback}: if we drop assumption SM1, the routine of a qualified call (such as q in the call \e{S.q (a)}, executed on behalf of \e{C1} in Fig. \ref{fig:oo_model}) may execute another qualified call with the same target (a call \e{C1.v (x)}).
Then the execution may find that object (here \e{C1}) in an inconsistent state.
The callback can be indirect (to a caller's caller and so on up the call chain). As to the practicality of this issue, recall that the 2016 \$50-million Ethereum cryptocurrency heist resulted from an exploited consistency violation in a callback \cite{Sirer_Ethereum_2016}.

\textbf{Furtive access}: while normally every routine \e{r} in a call \e{x.r (a)} must preserve the invariant, it is often necessary in practice, as an intermediate step, to allow a call to a utility routine that does not preserve it, violating SM2 and, as a consequence, breaking the Invariant Hypothesis.

\textbf{Reference leak}: if we drop SM3 (invariant locality), even if every operation on an object A preserves its invariant, an operation on \textit{another} object B, to which A holds a reference, may break it.

\subsection{The marriage example} \label{marriage_example}

A simple example, introduced in \cite{meyer_dependent_2005}, is subject to all three issues (other examples appear in section \ref{challenge_problems_and_solutions}).
It involves a single class \e{PERSON} and a single routine of interest, \e{marry}.
The class has attributes
\begin{lstlisting}[xleftmargin=3em]
spouse: detachable PERSON
is_married: BOOLEAN
\end{lstlisting}

\noindent where \e{spouse} is declared \e{detachable} as it can be a void (or ``null'') reference. It has the following invariant :
\begin{lstlisting}[xleftmargin=3em]
distinct: spouse $\neq$ Current
married_iff_has_spouse: is_married = (spouse $\neq$ Void)
reciprocal: is_married $\Rightarrow$ (spouse.spouse = Current) 
\end{lstlisting}
expressing the relationship between the attributes. ($\Rightarrow$ denotes implication, and a name before a colon in an assertion clause, such as \e{distinct}, is a tag identifying the clause.)
\e{is_married} could be a function (returning true if and only if \e{spouse $\neq$ Void}), but defining it as an attribute helps reveal potential problems.

\subsection{Recursive marriage and callbacks} \label{threats_to_sanity_clause:callbacks}

 \cite{meyer_dependent_2005} analyzes the difficulty of writing a procedure \e{marry (other: PERSON)} which will result in setting the \e{spouse} field of the current object to \e{other} and its \e{is_married} field to \e{True}.
The procedure has as a precondition that both the current object and \e{other} are unmarried.
A straightforward recursive implementation, where \e{marry} calls \e{other.marry (Current)}, will not work because this call will violate the second precondition.
The following variant avoids this problem:
\begin{lstlisting}[xleftmargin=1em]
marry_recursive (other: PERSON)
     require
         $\lnot$ is_married
         $\lnot$ other.is_married
         other $\neq$ Current
     do
        spouse := other
        if other.spouse = Void then --- This line causes a callback
           other.marry_recursive (Current)
        end
        is_married := True
      ensure
         spouse = other; other.spouse = Current
         is_married; other.is_married
      end
\end{lstlisting}
The call \e{other.spouse} is, as marked, a callback. It finds the original object inconsistent, violating \e{married_iff_has_spouse} since \e{spouse} is not \e{Void} but \e{is_married} is still \e{False} to satisfy the precondition.
No reshuffling of instructions will (to our knowledge) satisfy both the preconditions and the invariant.
All code is available in the article's repository \cite{supporting_material} for the reader to try out.

\subsection{Non-recursive marriage and furtive access} \label{threats_to_sanity_clause:slicing_model}

After callbacks, furtive access. In line with \cite{meyer_dependent_2005} it is possible to write a non-recursive version with auxiliary routines:
\begin{lstlisting}[xleftmargin=3em]
set_married
    do is_married := True end
set_spouse (other: PERSON)
    do spouse := other end
\end{lstlisting}
both visible, for obvious reasons of information hiding, only to \e{PERSON} (that is, not usable from other classes).
The callback goes away, but furtive access arises, in the form of violated invariants, for example if we write \e{marry} as follows (pre- and postconditions are as with the preceding version and omitted for brevity):
\begin{lstlisting}[xleftmargin=3em]
marry_stepwise (other: PERSON) -- Non-recursive version
  do
    set_married                 -- 1
    other.set_married           -- 2
    set_spouse (other)          -- 3
    other.set_spouse (Current)  -- 4
  end
\end{lstlisting}
Before instruction 4, the clause \e{married_iff_has_spouse} of \e{other}'s object invariant does not hold, violating the Invariant Hypothesis.
With the elementary semantics of invariants as suggested so far, such a violation will arise with any reordering of the four: as soon as one of instructions 2 and 4 has been executed, \e{other}  will violate its clause \e{married_iff_has_spouse} (and executing the other one of these instructions would be the only way to restore it).

The invariant at issue here is not the invariant of the current object but the invariant of \e{other}.
It is frustrating that these violations cause trouble, because they are all intuitively harmless: as in an internal (unqualified) call, the intermediate states of \e{other} can be inconsistent as long as the final one is consistent. In software verification, however, there is no simple equivalent of ``trust me, I know what I am doing, everything will be fine in the end!''. 

\subsection{Reference leak} 
\begin{figure}[h]
    \centering
    \includegraphics[scale=0.3]{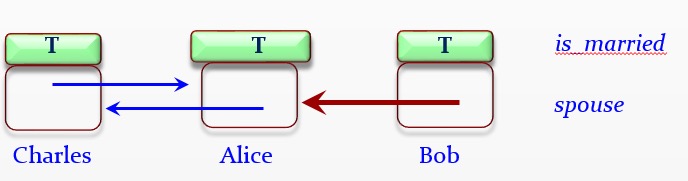}
    \caption{Incomplete remarriage}
    \label{fig:incomplete_remarriage}
\end{figure}

\noindent Finally, reference leak.
In the same example, assume that we somehow have obtained a solution for callbacks and furtive access and a suitable version of \e{marry}. A reference leak will occur, illustrated in Fig. \ref{fig:incomplete_remarriage}, if some code using references \e{Alice}, \e{Bob} and \e{Charles} to distinct objects performs
\begin{lstlisting}
     Alice.marry (Bob)
     Alice.divorce
     Alice.marry (Charles)
\end{lstlisting}
where \e{divorce} simply sets \e{is_married} to false and \e{spouse} to \e{Void} (its code appears in \ref{marriage_non_recursive}). Calling \e{divorce} is necessary because \e{marry}'s precondition.
After these calls, \e{Alice.spouse} is \e{Charles}, but \e{Bob.spouse} is still \e{Alice}, so \e{Bob.spouse.spouse} is \e{Charles}, causing \e{Bob} to violate \e{spouse.spouse} = \e{Current}.

The code is intuitively buggy: \e{divorce} should reset the \e{is_married} and \e{spouse} fields of the previous \e{spouse}.
But \e{divorce} satisfies its contract (with obvious pre- and postconditions), and in particular it preserves the class invariant for the target object, here \e{Alice}. 
The scary part is that one may break another object's invariant, here \e{Bob}'s, without performing any operation on it.
Any proof technique relying on a simple invariant concept (as in the Simple Model) will miss the bug as all routines satisfy their contracts.

Stating that the code is ``buggy'' means, more formally, that it violates the Invariant Hypothesis.
The violation will only cause trouble, however, in the \textit{next} qualified call, if any, of target \e{Bob}.
This property is the crux of the reference leak problem: it does not immediately manifest itself, but results in a corrupted data structure, with some sitting ducks --- inconsistent objects --- ready to fall prey to the next qualified call.

The rest of this article develops a sound proof rule addressing all three issues.

\section{Addressing callbacks: wrap up before calls} \label{callbacks}

We need a rule that accounts for callbacks.
Most published solutions require programmer annotations, which --- as noted --- limits their practicality.
Here the proof rule will integrate the sanitization of callbacks.
The GCP becomes stronger (before being weakened again in the next step, section \ref{slicing_model}).

\subsection{Wrapping up before going out} \label{wrapup}

The basic observation is that prior to executing a qualified call an object must make itself amenable to callbacks, meaning that it should guarantee its invariant.
Hence the next version of the rule:
\begin{figure}[H]
\centering
    \begin{tabular}{|cc|cc|}
         \cline{1-4}
         BS: & \es{INV$_S$} & AS: &\es{INV$_S$}  \\
         \hline
         \hline
         BC: & \colorbox{green!25}{\es{INV}\es{$_C$}} & AC: & \es{x.INV$_S$} \\
         \cline{1-4}
    \end{tabular}
    \caption*{\Rule{Callbacks}}
    \label{tab:callbacks_rule}
\end{figure}
\noindent For clarity, occurrences of {\e{INV}} in the statement of this rule refers to the corresponding class: \e{S} for the supplier side, \e{C} for the client side. 
The new element, highlighted in BC, is \e{INV$_C$}.
The use of \e{C} instead of \e{S} and of \e{INV} rather than \e{x.INV} is not a typo: we are talking of the invariant of the \textit{calling object}. Compare with the presence in \Rule{Pointless}, at the same position BC, of \e{x.INV}, meaning \e{x.INV$_S$}, imposing a Scandalous Obligation on the caller. Here, by using \e{INV$_C$}, the object takes care of ensuring its invariant before starting a qualified call (like a person who cleans up the house before going out on a walk, just in case someone comes to visit during that time).

The requirement of wrapping up before a qualified call will be loosened in section \ref{slicing_model}, and further in section \ref{smart} which  introduces a ``smart'' version of the invariant (removing the wrapping-up requirement if no callbacks may occur).

In some methodologies, as noted, ensuring the invariant before an outgoing call is the task of a special  \e{wrap} instruction  (section \ref{related_work}). The
\Rule{Callbacks} rule achieves the wrapping-up without imposing an annotation on  programmers.

\subsection{Computational model for handling callbacks} \label{callback_model}
The computational model corresponding to rule \Rule{Callbacks}, which we call the \textbf{Callback Model}, uses a new version \textbf{GCP-1} of Global Consistency.

\begin{figure}[h]
    \centering
    \includegraphics[scale=0.27]{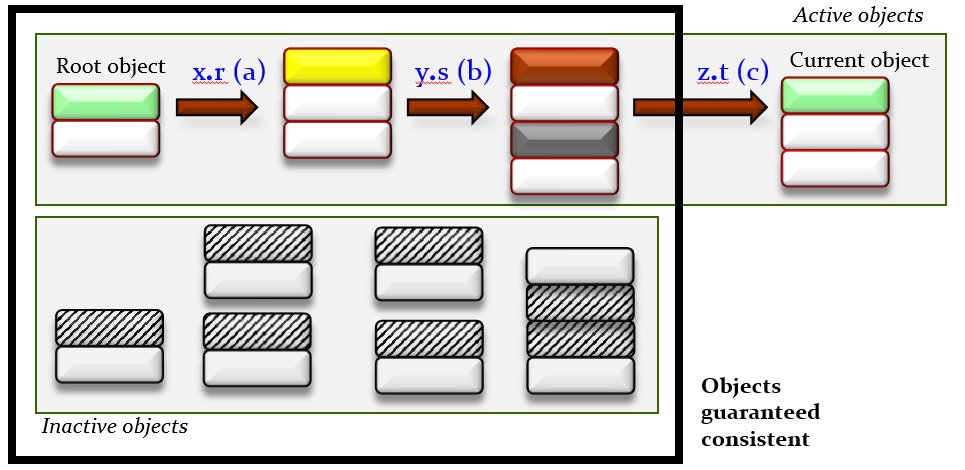}
    \caption{Object consistency in the Callback Model}
    \label{fig:snapshot_callback}
\end{figure}

Compared to GCP-0 from the Simple Model, GCP-1 restricts the set of objects that may violate their invariants, now limited to \textit{at most one} object, the current object. Fig. \ref{fig:snapshot_callback} (which drops inter-object references, not needed for the discussion) is the counterpart for GCP-1 to Fig. \ref{fig:snapshot_simple} for GCP-0. 

\noindent The Global Consistency Lemma now states that all operations, assuming SM2 and SM3 (but no longer SM1, invariant preservation) preserve GCP-1. As before (\ref{GCP-0}), each step demonstrates this property for one of the three types of event:
\begin{itemize}
    \item [G1]
       \textit{Internal operation}: same reasoning as for GCP-0: since we still assume invariant locality (SM3), the only object that may become inconsistent is the current object, which GCP-1 does not require to be consistent. 
    \item [G2]
    \textit{Start-of-call}: here the reasoning changes from GCP-0. We no longer assume SM1 (absence of callbacks) but now, remarkably, \textit{all objects will be consistent}, ensuring GCP-1: the previous current \e{C} thanks to precondition \e{INV}\e{$_C$} in entry \e{BC}; all other objects, which were already consistent by GCP-1. In particular, the target, which is the same as \e{C} in the case of a direct callback, is consistent.
    
   \item [G3]
   \textit{End-of-call}: the call's target \e{S} satisfies its invariant thanks to the postcondition \e{x.INV$_S$} in AC; all other objects satisfy theirs, so GCP-1 is preserved. Note that the previous current, which becomes current again, might be the same as \e{S} (direct callback) or another previously active object (indirect callback), and is consistent in all cases.  $\blacksquare$
\end{itemize}

\noindent In passing we have proved an extreme form of the GCP, the \textbf{Quasi-Constant Consistency Theorem} (QCCT-1): in the Callback Model, in all states other than immediately after an internal operation, \textit{all} objects are consistent. (Steps G2 and G3 of the proof showed that consistency holds of all objects after start-of-call and end-of-call, and internal events are the only other case. $\blacksquare$)

QCCT-1 is the closest we will ever come to a literal reading of the word ``invariant'' as a property that always (in this case, almost always) holds.

To prove the soundness of the Callback model for rule \Rule{Callbacks}, it suffices (see \ref{simple_model_sound}) to note that it satisfies the Invariant Hypothesis. That property is part of  QCCT-1: the target is consistent after start-of-call. (QCCT-1 also tells us that the target is consistent after end-of-call.) $\blacksquare$

\section{Addressing furtive access: the Slicing Model} \label{slicing_model}

Furtive access arises because of calls such as \e{other.set_married} (in non-recursive marriage, \ref{threats_to_sanity_clause:slicing_model}), where \e{other} does not satisfy its invariant, but where we do not expect it to; in fact, the call is expressly intended as a step towards re-establishing that invariant.
It is paradoxical and frustrating to be punished with an invariant violation just as we are working towards the invariant.

Although qualified, such calls are conceptually comparable to internal, unqualified calls \e{r (a)}, which can break the invariant of an object as part of a strategy to restore it.
Legitimate cases of furtive access arise when two or more objects are involved in a similar strategy, trying to restore each other's invariants through steps that may temporarily violate some of them.
The Observer example (sections \ref{challenge_problems_and_solutions:observer}, \ref{challenge_problems_and_solutions:subject}) is typical, involving a subject and observer objects.

The Slicing Model, which will be described now, is a small update to the Callback model, addressing furtive access.
The key observation for obtaining an annotation-free solution is that in practice legitimate cases of furtive access arise for calls to \textit{selectively exported} features.

\subsection{Selective export mechanisms} \label{selective_exports}

One of the object-oriented principles is information hiding (introduced by Parnas \cite{parnas_information_hiding} in the same year as the first article mentioning class invariants \cite{hoare_proof_1972}), which lets each class define which of its features and other properties it makes available (``exports'') to others (its clients). Many OO languages offer ways to fine-tune this decision through \textit{selective} export, whereby a class exports specific features to specific clients; examples are the ``friends'' mechanism of C++ and ``internal'' accessibility in C\#. In Eiffel, one may declare a feature in a clause starting with \e{feature \{A, B, $\ \text{\textellipsis}$\}} to specify that it is only exported (or ``visible'' per the terminology defined below in section \ref{slicing_terminology}) to classes \e{A}, \e{B}, \textellipsis\ and their descendants.

Previous work seems to have missed the relationship of selective exports to the furtive access problem of class invariants semantics, and its potential role in solving the problem without requiring additional programmer annotations.

\subsection{Taking advantage of selected exports} \label{using_selective_exports}

Observation of utility routines such as \e{set_married} and \e{set_spouse} and their counterparts in other furtive-access-prone code (such as the Observer pattern)  reveals that in properly written code they are not publicly exported. Instead, they are selectively exported to the relevant class, in this case \e{PERSON}  itself. As a consequence, while a call \e{Alice.marry\_stepwise (Bob)} (section \ref{threats_to_sanity_clause:slicing_model}) may appear in any class (since all versions of \e{marry} should be publicly exported),  \e{other.set\_married} is only permitted in class \e{PERSON} (and descendants).

Using information hiding criteria in the proof rule combines design principles and correctness concerns.
The reasoning is that a routine should not have to account for invariant clauses of visibility higher than its own.
All three invariant clauses of \e{PERSON} involve public features, \e{spouse} and \e{is_married}, of higher availability than \e{set\_spouse} and \e{set\_married}; they should not have to preserve these clauses.
The situation with examples such as Observer is similar.

\subsection{Slicing: definitions} \label{slicing_terminology}

As a reminder, an invariant is made of a sequence of clauses; the one for \e{PERSON}  (\ref{marriage_example}) has three clauses (here without tags):  \e{spouse} $\neq$ \e{Current},
\e{is_married} = \e{(spouse} $\neq$ \e{Void}) and 
\e{is_married} $\Rightarrow$ \e{(spouse.spouse = Current)}. The invariant is understood as the sequential conjunction (``and then'') of these clauses. 

Regardless of the language mechanism for selective exports, the \textbf{visibility \e{V (r)} of a feature \e{r}} of a class \e{C} is defined as the set of classes that may use \e{r} as clients, in qualified calls \e{x.r (a)} with \e{x} of type \e{C}.  The example has assumed so far (section \ref{threats_to_sanity_clause}) that the features of class \e{PERSON}  have full visibility (they are exported to all classes) except for two of them, exported only to class \e{PERSON} itself: \e{V (set_married)} = \e{V (set_spouse)} = \e{\{PERSON\}}. Visibility gives a preorder relation (reflexive and transitive) on the features of a class: we write \e{r} $\leq$ \e{g} to express that \e{V (r)} $\subseteq$ \e{V (g)} (\e{g} is exported to the same classes as \e{r}, and possibly to some additional ones).

The \textbf{queries of an invariant clause} are the queries that appear in it either by themselves or as targets of calls. For example, the queries of
 
    \e{a} \textbf{and} \e{b.c}
 
\noindent are \e{a} and \e{b} (but not \e{c}, which is the feature of a qualified call, not a feature of the invariant expression itself).

Visibility, defined above for individual features, also applies to invariant clauses. The \textbf{visibility \e{V (CL)} of an invariant clause \e{CL}} is the intersection $\cap$ \e{V (q)} of the visibilities of all its queries \e{q}. The preorder relation also follows: \e{CL1} $\leq$ \e{CL2} means that $\cap$ \e{V (q1)} $\subseteq$ $\cap$ \e{V (q2)} for queries \e{q1} of \e{CL1} and \e{q2} of \e{CL2}; in other words, for any class \e{C} to which all queries of \e{CL1} is exported, all queries of \e{CL2} are also exported to \e{C}. The preorder relation can similarly hold between invariant clauses and features: \e{CL} $\leq$ \e{r} means that \e{V (CL)} $\subseteq$ \e{V (r)}, that is to say, \e{r} is exported to all classes to which all queries of \e{CL} are exported.

Finally, if \e{INV} is the invariant of a class, its \textbf{slicing} by  a feature \e{r} of that class, written \e{INV / r}, is \e{INV} restricted to its clauses CL such that \e{CL} $\leq$ \e{r}. In other words, \e{INV / r} is the part of \e{INV} that \textbf{has no move visibility} than \e{r}.

The clauses of the invariant of class \e{PERSON} (repeated at the beginning of this subsection) all have \e{spouse} or \e{is_married} among their features, and hence have full visibility since these features are public (exported to all classes). Since \e{marry}, in all its versions, is also public, \e{INV / marry} is the full \e{INV}. On the other hand, \e{INV / set_spouse} and \e{INV / set_married} are empty invariants since the given features have lesser visibility (they are exported to \e{PERSON}  only).
In the examples summarized in section \ref{challenge_problems_and_solutions}, sliced invariants are in-between these extremes.

The following \textbf{Slicing Implication Theorem} holds: if \e{r} $\leq$  \e{g}, then \e{INV / g} $\Rightarrow$ \e{INV / r}.
Proof: for any clause \e{CL} of \e{INV / r}, \e{CL} $\leq$ \e{r} by the definition of slicing. Hence \e{CL} $\leq$ \e{g},  implying --- again by the definition of  slicing, applied now to \e{INV / g} --- that \e{CL} is one of the clauses of \e{INV / g}. So all the clauses of \e{INV / r} are also clauses of \e{INV / g}. As a consequence, remembering that an invariant denotes the conjunction of its clauses, \e{INV / g} implies \e{INV / r}. $\blacksquare$
 
A caveat about  slicing is that some invariant clauses may need reinforcing to remain defined. For example, a good prover will accept an invariant clause \e{c}  involving \e {1 / y} if earlier clauses state \e{x = y} and \e{x > 0}; slicing \e{x} off removes these clauses and makes \e{c} potentially undefined, hence useless for verification. It is the programmer's responsibility to make sure that all invariant clauses are meaningful, pre- and post-slicing (here, precede \e{c}'s condition by \e{y} \e{$\neq$} 0 \e{$\Rightarrow$}).

\subsection{Extended-active objects} \label{extended_active}

\begin{figure}[h]
    \centering
    \includegraphics[scale=0.27]{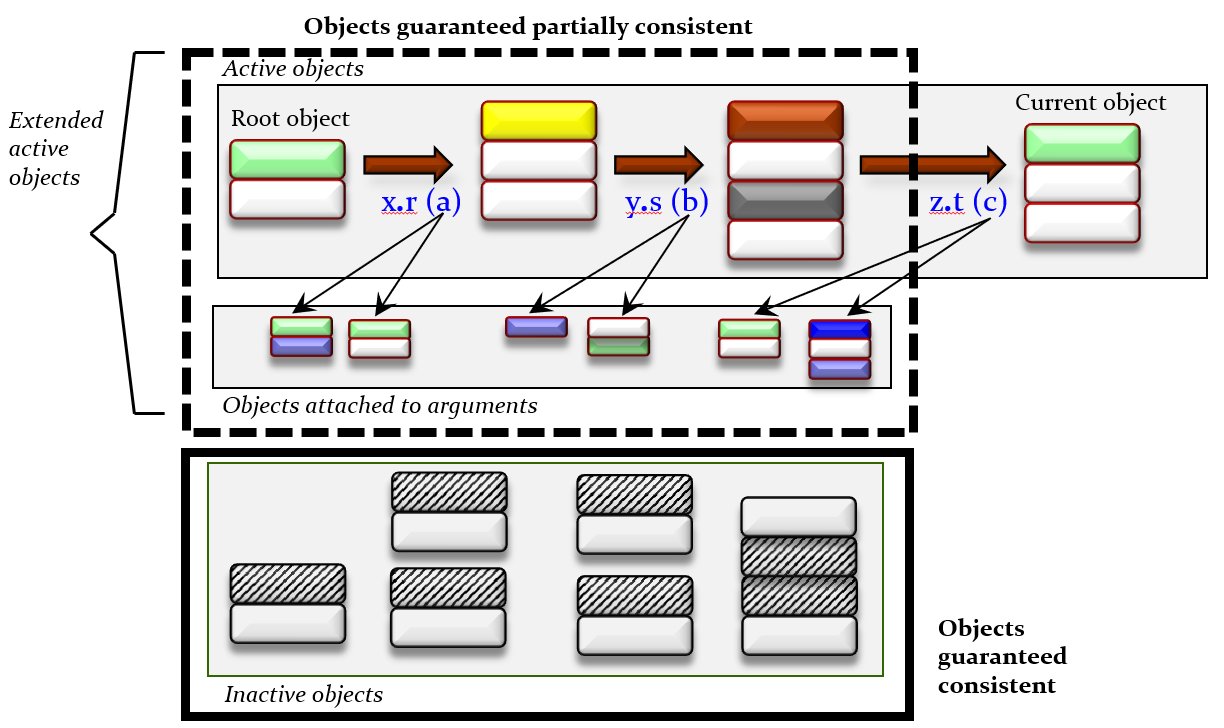}
    \caption{Object consistency in the Slicing Model}
    \label{fig:snapshot_slicing}
\end{figure}

Active objects have been defined as the objects in the call chain (Figs. \ref{fig:snapshot_simple}, \ref{fig:snapshot_callback}). Each is executing ``its call'' to ``its routine'', \e{r, s, t} in the figures. These routines can have arguments of reference types. The term \textbf{extended-active objects} will cover active objects  plus those attached to arguments of calls in the call chain. (Remember that an object is ``attached to'' a reference if that reference points to it.) For brevity, ``inactive'' will from now on denote all \textbf{non-extended-active} objects (in the thick rectangle at the bottom of Fig. \ref{fig:snapshot_slicing}). Note that the current object is active and hence extended-active, but all the properties below involving extended-active objects will be restricted to non-current ones.

As illustrated in Fig. \ref{fig:snapshot_slicing} and explained next, in the Slicing Model all non-current extended-active objects (those in the large dashed rectangle) will satisfy their invariants sliced by the routine of the respective call. 

\subsection{Relative consistency and the Slicing Model} \label{relative_consistency}

Consider a non-current extended-active object, with invariant \e{INV}. It either started the execution of a qualified call \e{x.r (a)} or is one of its arguments \e{a}.  It is \textbf{partially consistent} if it satisfies \e{INV / r}: the invariant sliced by the current routine. Obviously, consistent implies partially consistent. A non-current object is \textbf{relatively consistent} if it is either consistent or partially consistent .

GCP-2, the new variant of the Global Consistency Property (Fig. \ref{fig:snapshot_slicing}), differs only slightly from GCP-1 (compare to Fig. \ref{fig:snapshot_callback}): \textbf{all non-current objects are relatively consistent}. In particular, any (non-current) object in the call chain or attached to an argument is partially consistent (it satisfies \e{INV / r} for its routine \e{r}). All other non-current objects are fully consistent.

The Invariant Hypothesis (\ref{invariant_hypothesis}) similarly becomes the Relative Invariant Hypothesis, stating that \textbf{at the start of any call, the target satisfies its relative invariant} (its invariant sliced by the feature of the call).

\subsection{The slicing rule} \label{slicing_rule}

\textit{Notation}: \e{L.INV}, for a list \e{L},  is the conjunction of \e{x.INV} for all elements \e{x} of \e{L} (in the rule, L is a list of formal or actual arguments, \e{f} or \e{a}). Here is the rule:
\begin{figure}[H]
\centering
    \begin{tabular}{|cc|cc|}
         \cline{1-4}
         BS$_1$: & {\es{INV}}\colorbox{green!25}{ / \es{r}} & AS$_1$: &\es{INV} \colorbox{green!25}{\es{/ r}} \\
         BS$_2$: & \colorbox{green!25}{\es{f.}\es{INV}}\colorbox{green!25}{ / \es{r}} & AS$_2$: &\colorbox{green!25}{\es{f.INV}}  \\

         \hline
         \hline
         BC$_1$: & \es{INV}\colorbox{green!25}{ / \es{r}} & AC$_1$: & \es{x.INV} \colorbox{green!25}{\es{/ r}} \\
         BC$_2$: & \colorbox{green!25}{\es{a.}\es{INV}}\colorbox{green!25}{ / \es{r}} & AC$_2$: & \colorbox{green!25}{\es{a.INV}} \\
         \cline{1-4}
    \end{tabular}
    \caption*{\Rule{Sliced}}
    \label{tab:sliced_rule}
\end{figure}

\noindent  (We no longer need \e{C} and \e{S} subscripts for \e{INV}.) Notes on rule \Rule{Sliced}:
\begin{itemize}
\item
Preconditions on both the target (BS$_1$, BC$_1$) and arguments  (BS$_2$, BC$_2$) are both sliced.

\item

Postconditions are sliced on the target (AS$_1$, AC$_1$), indicating that the routine need only preserve the relevant part of the invariant, to handle furtive access. An argument, however, denotes an object that may become inactive after the call, rejoining the lower part of Fig. \ref{fig:snapshot_slicing}, and hence must be fully consistent on routine exit, ready to serve later as the target of a call to a routine of arbitrary visibility. The proof will formalize this analysis.     

\item
The Slicing Implication Theorem (\ref{slicing_terminology}) indicates that some preconditions (BS, BC) are weaker than their counterparts in  \Rule{Callbacks}. The effect for a selectively exported routine is to make the proof of the routine's correctness harder (it assumes less) and the proof of a particular call easier (the object only needs to wrap itself up for its sliced invariant, not the full one). This property is what makes it possible to handle tricky cases such as Observer or marriage without annotations.

\end{itemize}

\subsection{Two programming language conditions} \label{slicing_condition}
    
The soundness of the \Rule{Sliced} rule requires two conditions on programs, to be enforced by the programming language.

\textbf{Export Consistency Condition}: in any qualified call \e{x.s (a)} appearing in a routine \e{r}, \e{s} $\leq$ \e{r}.

\begin{figure}[h]
                \centering
                 \includegraphics[scale=0.27]{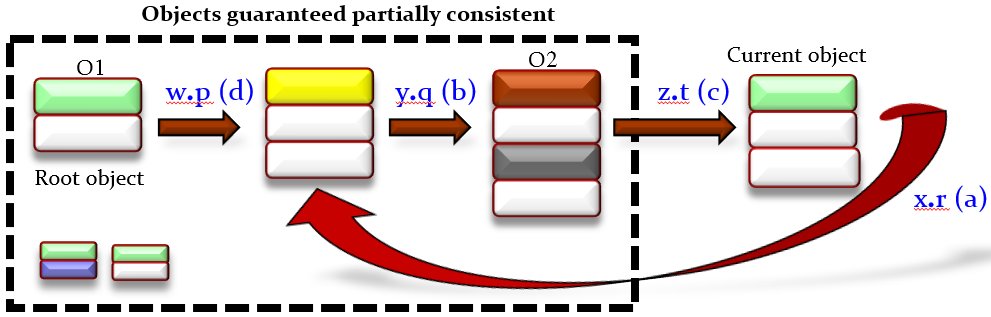}
                 \caption{A call chain with decreasing visibility and an indirect callback}
          \label{fig:callback_to_active}
        \end{figure}

In other words, a routine may only perform qualified calls to routines with the same visibility or less. In the example of Fig. \ref{fig:callback_to_active}, visibility along the routine sequence \e{p, q, t, r} can only remain the same or decrease. This reasonableness rule (do not rise above your own rank) can be compared to the practice of Web browsers such as Firefox, which let users start a new public or private window from a public window, but, from a private window, only a private one.

\noindent  Under the Export Consistency Condition, the following \textbf{Safe Indirect Callback Theorem} holds: in an indirect callback of routine \e{r}, the routine \e{q} of the target has the same visibility as \e{r}. (See Fig. \ref{fig:callback_to_active}.) \textit{Proof}: from the Export Consistency Condition, \e{r} $\leq$ \e{q} and \e{q} $\leq$ \e{r}. (From \ref{slicing_terminology}, ``$\leq$'' on features is only a preorder, but on the associated visibility sets such as \e{V (r)} the relation is antisymmetric since it is simply ``$\subseteq$''.)

The second language condition needed for the soundness of the Slicing Model is the \textbf{Selective Export Call Rule}: in a call \e{x.r (a)} where \e{r} is selectively (rather than fully) exported, the target \e{x} must be a formal argument of the enclosing routine. (In other words, \e{x} may not be an attribute or a local variable.) This restriction does not affect the actual power of the programming language because it is always possible to wrap a call in a routine. Informal examination of code using selective exports shows that most selective-export calls are indeed on arguments and that only a small amount of tweaking may be needed, as in two of the challenge-problem examples, marriage (\ref{marriage_example}) and circular list (\ref{circular_list}), whose verification led to introducing an extra argument.

The soundness of the Slicing Model also assumes that a routine cannot change the values of reference variables passed to it as actual arguments in a call. In other words, such arguments are passed by value. Many modern OO languages (such as Eiffel and Java) satisfy this common-sense condition, but others may violate it, requiring adaptations of the model presented here.

\subsection{Handling furtive access correctly: the slicing rule is sound} \label{soundness_furtive}

As with earlier rules, the core of the soundness proof for \Rule{Sliced} is the proof of the Global Consistency Lemma, here stating the preservation of GCP-2 (all objects except the current one are relatively consistent) through events of all three kinds:

\begin{itemize}
   
    \item [G1]
            \textit{Internal operation}: the reasoning for \Rule{Callbacks} (GCP-1) applies: since internal operations only affect the current object, and we still assume SM3 (invariant locality), they preserve GCP-2. Note the importance of the rule that routines cannot change the value of their reference arguments (\ref{slicing_condition}): without it, an assignment \es{f := Current} to a formal argument \e{f}, modifying the corresponding actual, could make an extended-active object inconsistent.
    \item [G2]
            \textit{Start-of-call}: the following objects are involved:
            
           \begin{itemize}
               \item 
            (G2-A) The target of the call. It becomes current and hence does not affect GCP-2.
   
            \item 
            (G2-B) The client (the initiator of the call), previously current. It remains active and, thanks to the call's precondition \es{INV} \es{/ r}  (entry BC$_1$), satisfies the sliced invariant, as required of active objects under GCP-2.
            
                  \item 
            (G2-C)
            Objects attached to arguments of the call (\e{a} in the caller, \e{f} in the target). They become extended-active and hence must satisfy the invariant sliced by the routine being called. That is exactly what the precondition AS$_2$ and its counterpart BC$_2$ enforce: \es{a.}\es{INV} \es{ / r}.
            
            \end{itemize} 
        \item [G3]
      \textit{End-of-call}: the following objects are involved:
      \begin{itemize}
          \item 
          (G3-A)
          The target. It ceases to be current. From entry AS$_1$ it satisfies \e{INV / r}. If (outside of the case of a callback, in which it remains active) it could become inactive, that would not be enough, since inactive objects must satisfy the \textit{full} invariant under GCP-2. The Selective Export Call Rule (\ref{slicing_condition}) saves us: by requiring the target of a selective-export call to be a formal argument of the calling routine, it guarantees that the target remains extended-active (attached to an argument of an active routine), for which the sliced invariant suffices.
          
          This property is key to the soundness of the Slicing Model, enabling ``furtive access'' calls to use the sliced invariant on both entry and return. The proofs of such challenge examples as marriage and Observer (\ref{circular_list}) would be impossible (and the software incorrect) if furtive-access routines had to restore the full invariant.
          \item
         (G3-B)
         The client. It becomes current again, hence incurs no obligation.
          \item 
        (G3-C)
        Objects attached to arguments of the call. During the call they were extended-active and only had to be partially consistent. On end-of-call they cease to be extended-active (except if active or attached to an argument higher up in the call chain) and must become \textit{fully} consistent to preserve GCP-2. That is exactly what the postconditions AS$2$ and its counterpart AC$_2$ guarantee: \e{a.INV}, using the unsliced \e{INV}.
  $\blacksquare$
            \end{itemize}

 \end{itemize}
  
  \noindent Steps G2-A and G3-B involve an object that GCP-2 allows to be inconsistent because it is current. It is in fact relatively consistent in both cases, yielding an updated form QCCT-2 of the \textbf{Quasi-Constant Consistency Theorem}: in the Slicing Model, in all states other than immediately after an internal operation, \textit{all} objects are relatively consistent. \textit{Proof}:
  \begin{itemize}
      \item 
  In start-of-call (G2-A), the new current object is the target, which may be: inactive, satisfying the full \e{INV}; attached to an argument, hence relatively consistent by precondition BC$_2$; in the case of a direct callback, the previous current object, which is relatively consistent thanks the ``self-wrapping-up'' precondition BS$_1$; or (in the remaining case, indirect callback) an object up the call chain --- active, not just extended-active --- which is relatively consistent as a consequence of the Safe Indirect Callback Theorem (\ref{slicing_condition}).
  
    \item
    In end-of-call (G3-B), the new current object is the client, which becomes active and hence, by preservation of GCP-2, relatively consistent. $\blacksquare$ 
  
  \end{itemize}

\noindent As a consequence of QCCT-2, the Slicing Model satisfies the Invariant Hypothesis and hence (see \ref{simple_model_sound}) is sound. $\blacksquare$

\section{Programming style and smart invariants} \label{smart}

The rules seen so far place two limitations on programmers: the obligation for an object to wrap itself up before a qualified call (section \ref{callbacks}) and the prohibition of qualified calls to routines of greater visibility (Selective Export Call Rule, \ref{selective_exports}). They may appear unduly constraining for practical programming. It is possible to relax them considerably, removing much of the burden on programmers.

Both of the restrictions cited arise from the need to protect against callbacks. 
In practice, many calls \e{x.r (a)} are \textbf{callback-free}, written \e{CBF (r)}, meaning that \e{r} and any routine it calls directly or indirectly do not call back into the current object. For example, \e{r} may be a routine from a reusable library (as in \e{x.extend (a)} which adds an element \e{a} to the end of a list \e{x}), written long before the client code and hence certain not to call it. (The present work  assumes that no arguments are function pointers or their counterparts such as C\# delegates and Eiffel agents.) It may be possible for the verifying tools to ascertain \e{CBF (r)} in such cases through modular analysis (by making sure that any compiled routine carries information about which routines it may call outside of its own package).

If it can be ascertained that a call is callback-free call the rules simplify:
\begin{itemize}
    \item 
    Objects no longer need to wrap themselves up before a qualified call. 
    \item
    The  Selective Export Call Rule is no longer required.
\end{itemize}

\noindent Under these changes, the proof of the soundness of the Sliced Model \ref{soundness_furtive}) remains applicable with only two updates:
\begin{itemize}
    \item 
    Partially relax the Global Consistency Property (GCP-2): remove the requirement of partial consistency on objects (such as the first three at the top of figure \ref{fig:snapshot_slicing}) in a call chain that is guaranteed callback-free. Like the current object, they are not subject to any conditions. Requirements on objects attached to \textit{arguments} remain.
    
        \item
        
        Step G2-B showed that the client, which ``remains active'', still ``satisfies the sliced invariant'' thanks to the call's precondition. That precondition may not hold any more, but we no longer need active objects to satisfy the sliced invariant, since the only reason was to handle callbacks correctly.
        
\end{itemize}

\noindent Everything else remains. To reflect this change in the rule, it suffices to make a small adaptation to the definition of slicing. Consider \e{INV / r}, the local  invariant \e{INV} sliced by routine \e{r}. (We call \e{INV} ``local'' because it is applied to the current object, to distinguish it from the ``remote'' invariant \e{x.INV}, handled next.) Let \underline{\e{INV / r}} be the sliced local invariant as defined earlier (section \ref{slicing_terminology}). Thanks to the preceding observations the rules involving the sliced local invariant remain sound if we replace it by the \textbf{smart sliced local invariant}, defined as  

\begin{center}
    CBF (r) \e{$\lor$} \underline{\e{INV /r}}     
\end{center}

\noindent In other words, the usual invariant by default, but True if the verification tools can guarantee that the call will not cause a callback. Using the smart local invariant is only an optimization: wherever the tools cannot guarantee \e{CBF (r)}, the standard sliced local invariant remains applicable.

With the smart local invariant and the loosened Selective Export Call Rule, the impact on programming style is small. Where a constraint remains, because calls cannot be guaranteed callback-free, it is easy to justify it by the potentially catastrophic consequences of an incorrect callback, evidenced by the DAO hack.

We may apply a similar small optimization to \textit{remote} sliced invariants applied to arguments, such as \e{a.INV / r} in the caller-side precondition BC$_2$ of \Rule{Sliced}. That condition is only necessary for arguments \e{a} that are neither current nor active, a condition we write \e{NCA (a)}. Let \underline{\e{a.INV / r}} be the sliced remote invariant as defined earlier (\ref{slicing_terminology}). The rules involving the sliced remote invariant remain sound if we replace it by the \textbf{smart sliced remote invariant}, defined as

\begin{center}
     \ \ \ \ \ NCA (a) \e{$\lor$} (\underline{\es{a.}\es{INV / r}})   
\end{center}

\noindent In practice, NCA may be more delicate to ascertain than CBF, leading to using plain \es{a.}\es{INV}. The smart version highlights a particular difficulty of calls, familiar to anyone having tried to verify or just debug OO programs: passing to another object a reference to oneself (\e{Current}, ``\e{this}'') reeks of trouble. The preconditions ensure the key property (giving some insight into conditions governing the practice of OO programming): you may only pass yourself as argument if you are sure you are consistent --- and so are other active objects up the call chain. 

The verification of the challenge examples reported in section \ref{challenge_problems_and_solutions} used the plain invariants, not the ``smart'' versions.

\section{Addressing reference leaks} \label{reference_leaks}

Rule \Rule{Sliced} dropped assumption SM1 (no callback) and relaxed SM2 (invariant preservation) from the Simple Model.
There remains to remove SM3, invariant locality, stating that the invariant only involves the current object's fields.

Soundness proofs so far have relied on this assumption by considering that the only way to invalidate an object's invariant is by modifying one of its fields; of the three kinds of event (\ref{simple_model_sound}), only ``internal operations'' can do this.
Without invariant locality, such an operation can also invalidate the invariant of other objects which \textbf{depend} on it.
Considering dependents leads to new versions of the rule, ``strong'' and ``weak''.
They differ in terms of both implementation (in a verifier) and underlying effect on consistency:
\begin{itemize}
    \item The strong version is more precise but (in the absence of annotations) requires global analysis, whereas all the previous rules are modular (meaning that they can be verified class by class).
    The weak version, in its two variants, is modular but less powerful.
    \item The strong version retains GCP-2 (all non-current objects are relatively consistent).
    Operations can therefore assume the invariant (possibly sliced) on entry.
    With the weak versions, some dependent objects satisfy only its ``local'' part; calls on these objects can no longer expect the full invariant.
    \item The weak versions, then, have a ``lazy'' approach to consistency: the object structure may contain inconsistent objects.
    The inconsistency will be detected (in static or dynamic verification) only in case of a later attempt to use them.
    Absent such calls, it will remain.
    The approach remains sound: all objects actually used satisfy their (relative) invariants.
    It is a matter for discussion (in the spirit of ``\textit{if a tree falls unheard and unseen in the forest}'') whether inconsistencies in unused objects are a reason to worry.
\end{itemize}

\subsection{Strong rule} \label{reference_leaks:strong}
We need a precise definition of the notion of dependents of an object.
At any time during execution, the dependent set of an object \e{S}, \e{Dep (S)}, is the set of other objects \e{C} whose invariants have a clause \e{x.some_property} where \e{x} is a field of \e{C} attached to \e{S}.
In Fig. \ref{fig:incomplete_remarriage}, for example, \e{Dep (Alice) = \{Bob, Charles\}}.

In reference to the program text, we may use \e{Dep (x)} to denote \e{Dep (S)} for the object \e{S}  attached to \e{x} at run time (empty if \e{x} is \e{Void}).
If we could determine that set, the corresponding \Rule{Strong} proof rule would be the same as \Rule{Sliced} with a clause on dependents added above and below (highlighted).
\begin{figure}[H]
\centering
\setlength\extrarowheight{-2pt}
    \begin{tabular}{|cc|cc|}
         \cline{1-4}
         BS$_1$: & \es{INV / r}                & AS$_1$: & \es{INV / r}              \\
         BS$_2$: & {\es{f.}\es{INV}}{ / \es{r}} & AS$_2$: &{\es{f.INV}}  \\

         BS$_3$: & \cellcolor{green!25}\es{Dep (Current).INV / r}  & AS$_3$: & \cellcolor{green!25}\es{Dep (Current).INV / r} \\
         \hline
         \hline
         BC$_1$: & \es{INV} \es{/ r}                & AC$_1$: & \es{x.INV / r}                   \\
        BC$_2$: & {\es{a.}\es{INV}}{ / \es{r}} & AC$_2$: & {\es{a.INV}} \\
        BC$_3$ &      \cellcolor{green!25}\es{Dep (Current).INV / r}      & AC$_3$: & \cellcolor{green!25}\es{Dep (x).INV / r}   \\

         \cline{1-4}
    \end{tabular}
    \caption*{\Rule{Strong}}
    \label{tab:strong_rule_leaks}
\end{figure}
\noindent Soundness: GCP-2 remains.
So does the previous proof of soundness (\ref{soundness_furtive}), with Invariant
Locality relaxed into ``an object's invariant only involves fields of the object
itself and its dependents''. $\blacksquare$

In effect, the \Rule{Strong} rule addresses -- rather than the classical notion of class invariant -- the notion of multi-object invariant (\ref{terminology}) by extending the invariant of an object with the invariants of all its dependents.  Implementing this rule into a verifying tool requires computing the \e{Dep} sets, which in the absence of programmer annotations implies an analysis that is global (on the entire program) rather than modular (class by class). As a conjecture, a modular implementation may be applicable to classes whose invariants satisfy specific patterns (see the ``mirroring'' suggestion in section 11.14 of \cite{oosc_2}).

\subsection{Weak rule, call-side} \label{reference_leaks:weak_client}
\Rule{Weak-C} (``C'' for ``Client'' or ``caller'') puts the burden on callers:
\begin{figure}[H]
\centering
\setlength\extrarowheight{-2pt}
    \begin{tabular}{|cc|cc|}
         \cline{1-4}
         BS$_1$: & \es{INV} \es{/ r}            & AS$_1$: & \es{INV / r }                             \\
        BS$_2$: & \es{f.}\es{INV}{ / \es{r}} & AS$_2$: &{\es{f.INV}}  \\
         \hline
         \hline
         BC$_1$: & \es{INV / r}            & AC$_1$: & \es{x.INV / r }          \\
         BC$_2$: & {\es{a.}\es{INV}}{ / \es{r}} & AC$_2$: & {\es{a.INV}} \\
         BC$_3$: & \cellcolor{green!25}\es{x.INV\_REM / r}     &             &    \\
         \cline{1-4}
    \end{tabular}
    \caption*{\Rule{Weak-C}}
    \label{tab:weak_c}
\end{figure}
\noindent This rule relies on splitting any invariant into two parts: \e{INV_LOC} and \e{INV_REM}, for ``local'' and ``remote''.
\e{INV_REM} contains the clauses that have a subexpression of the form \e{b.some_property} where \e{b} denotes a reference field.
In \e{PERSON} (\ref{marriage_example}), \e{INV_LOC} includes two of the three clauses \e{distinct} and
\e{married_iff_has_spouse}; the third one, \e{reciprocal}: \e{is_married $\:\Rightarrow\:$ (spouse.spouse = Current)}, goes into  \e{INV_REM} (because of \e{spouse.spouse}).

We say that an object is \textbf{locally consistent} if it satisfies the local part of its invariant. (Local consistency could also be partial in the sense of slicing as defined in \ref{relative_consistency}, but will only be applied in the unsliced case.)

The weak variant is based on the observation that the Invariant Hypothesis would hold if the invariant consisted of \e{INV_LOC} only; the risk of reference leak comes from \e{INV_REM}, which GCP-2 does not allow us to assume on start-of-call. \textbf{GCP-3}, the new GCP, is (change from GCP-2 highlighted):
\begin{itemize}
    \item All non-current active objects are relatively consistent.
    \item All inactive objects are \colorbox{green!25}{locally} consistent (with the unsliced invariant).
\end{itemize}

\noindent The inactive part of the heap (bottom part of Fig. \ref{fig:snapshot_slicing} and predecessors) consists of objects that are ``just sitting there'', ready to serve as target of future calls. In GCP-2 and previous models, they were fully consistent, in anticipation of ensuring the Invariant Hypothesis in those calls. With GCP-3 this massive guarantee of consistency no longer holds: we must cope with the possible presence of inactive objects that are only \textit{locally} consistent. Under \Rule{Weak-C}, callers can only rely on the local consistency of these objects, and must ensure the rest.

Soundness: the Invariant Hypothesis continues to hold for the local part of the invariant, and for the remote part is ensured by the caller (entry BC$_3$). $\blacksquare$ 

Rule \Rule{Weak-C}, by reintroducing a limited form of Scandalous Obligation (a partial return to \Rule{Pointless} of section  \ref{sound_useless}), makes the client responsible for invariant elements that could have been messed up by a ``man in the middle'' such as Bob in the marriage case.
The next rule frees the client from the Scandalous Obligation by making life harder for the \textit{supplier} instead.

\subsection{Weak rule, supply-side} \label{reference_leaks:weak_supply}

In rules governing routine calls, we may always move an obligation from the client's to the supplier's precondition.
Rule \Rule{Weak-S} (``S'' for ``Supplier'') transfers responsibility for the remote part of the invariant from the client (as in \Rule{Weak-C}) to the supplier. Removing \e{INV_REM} on one side means adding \e{INV_LOC} to the other, maintaining the Invariant Hypothesis and hence soundness:

\begin{figure}[H]
\centering
\setlength\extrarowheight{-2pt}
    \begin{tabular}{|cc|cc|}
         \hhline{-|-|-|-}
         BS$_1$: & \colorbox{green!25}{\es{INV_LOC}} \es{/ r}       & AS$_1$: & \es{INV / r} \\
         BS$_2$: & {\es{f.}\es{INV}}{ / \es{r}} & AS$_2$: &{\es{f.INV}}  \\
         \hline
         \hline
         BC$_1$: & \es{INV} \es{/ r}            & AC$_1$: & \es{x.INV / r}                      \\
        BC$_2$: & {\es{a.}\es{INV}}{ / \es{r}} & AC$_2$: & {\es{a.INV}} \\
         \cline{1-4}
    \end{tabular}
    \caption*{\Rule{Weak-S}}
    \label{table:weak_s}
\end{figure}
\noindent A criticism of this rule is that it hands over some of the responsibility from the verifier to the programmer.
For each routine:
\begin{itemize}
    \item It is not possible to assume \e{INV_REM} on entry (only \e{INV_LOC}).
    That situation seems inevitable in the absence of an analysis that would compute (or approximate) the \e{Dep} sets.
    \item As a consequence, the routine may have to do more work, since it is still required to satisfy the full invariant on exit (AS).
    \item Note that there is no change to the precondition on the caller side (BC). We make it harder for the routine, forcing it to handle objects that do not satisfy their multi-object requirements (since the remote part of the invariant does not hold), but there is no corresponding extra work for the caller, and nothing the caller can do to help.
\end{itemize}

\noindent The approach is ``lazy'', as noted, in that it may leave inconsistent objects around.
This possibility causes no threat to program correctness since subsequent operations on such an object will have to restore consistency.
With the strong approach such a situation would not arise, but the price to pay is that operations on an object must also take care of updating other objects whose consistency depends on it.

While \Rule{Strong}  remains a possible choice, \Rule{Weak-S} appears, in spite of the preceding objections, to be the best available rule for proving the correctness of object-oriented programs and the main result of the present work. It was given in the usual style of Hoare inference rules (without the abbreviations introduced in the last sections) in the Preview section (\ref{progression}) and is used for the  proofs of example programs in section \ref{challenge_problems_and_solutions}. 

\section{Initialization, queries and concurrency} \label{initialization_queries_concurrency}
The discussion so far has addressed commands (procedures): operations that can modify an existing object, causing transitions between consistent states S$_i$ of its lifecycle as illustrated in Figs \ref{fig:object_lifecycle} and \ref{fig:oo_model}.
We also need rules for two separate cases: a creation operation, which does not assume an existing object but creates a new one; and a query, which instead of modifying an existing object, returns information about it. The following discussion will also mention the effect of concurrency and summarize the applicable rules.

\subsection{Initialization} \label{initialization_queries_concurrency:initialization}

\Rule{Weak-S} and predecessors govern a qualified call \e{x.r (a)} where \e{x} denotes an existing object which, per the Invariant Hypothesis, satisfies the invariant (at least its local component).
Such rules correspond to the induction step in a proof by induction.
We also need an initialization rule, corresponding to the base step.

The initialization mechanism is part of object creation, variously written: 
\begin{itemize}
\item  \e{create x.make (a)} in Eiffel, where \e{make} is a creation procedure.
\item \e{x = new SC (a)} in Java, C\# and C++, for a class \e{SC} with an appropriate ``constructor'' determined by the types of arguments in \e{a}.
\end{itemize}

\noindent The goal of object creation is to obtain an object satisfying its invariant.
In adapting the preceding rules, it suffices to remove any property of the object itself on the precondition
side --- \e{INV} for the supplier (BS) and \e{x.INV} for the client side (BC) which makes no sense for a creation instruction as it creates a new object from scratch. (Of course any specific precondition of the creation procedure is still applicable; remember that the tabular notation used since \ref{ideal_rule} ignores such routine-specific pre- and postconditions, but they reappear the rule  when we express it in full, in the classical notation of Hoare logic used for example for \Rule{Weak-S} in \ref{progression}.) The postcondition side remains, as well as the conditions on arguments. The creation variant of \Rule{Weak-S}, applicable to an instruction \e{create x.make (a)}, is \Rule{Weak-S-Creation}:
\begin{figure}[H]
\centering
\setlength\extrarowheight{-2pt}
    \begin{tabular}{|cc|cc|}
         \cline{1-4}
         BS$_1$: & \colorbox{green!25}{\es{DEF$_S$}}         & AS$_1$: & \es{INV}      \\
        BS$_2$: & {\es{f.}\es{INV}}& AS$_2$: &{\es{f.INV}}  \\

         \hline
         \hline
         BC$_1$:  & \cellcolor{green!25}\es{\ }  & AC$_1$: & \es{x.INV}                              \\
         BC$_2$: & {\es{a.}\es{INV}} & AC$_2$: & {\es{a.INV}} \\

         \cline{1-4}
    \end{tabular}
    \caption*{\Rule{Weak-S-Creation}}
    \label{tab:weak_s_creation_rule}
\end{figure}
\noindent DEF$_S$ denotes the property that fields of the new object have default values (typically zero for integers etc.). Including DEF$_S$ is appropriate for languages such as Eiffel, Java and C\# which specify default initialization rules; for C++, which does not, the BS$_1$ entry remains blank. Since in a creation instruction there is no prior ``current'', the rule needs neither slicing nor (as a consequence) the ``smart'' versions of the invariants (section \ref{smart}): a just created object must always satisfy the full invariant.

\subsection{Pure queries} \label{initialization_queries_concurrency:pure_queries}
A query is an operation returning information about an object.
It can be an attribute, giving the value of a field of the object, or a function, returning the result of a computation on those fields.

It is common programming practice (violating the ``\textit{Command-Query Separation}'' design principle,  \cite{oosc_1}, \cite{oosc_2}) to let functions change objects.
Then the above rules for procedures are applicable to functions, with postconditions extended by properties of function results.
The question remains of \textit{pure} queries, which cannot change the state, and include attributes and side-effect-free functions.

The postcondition side of  previous rules, AS and AC, is irrelevant for a routine that cannot change any object. 
Pure-query rules correspondingly have only one column (BS and BC), the same as in the corresponding rules for procedures.

The results of this simplification (with no new elements, just a removal of the postcondition column in the general rules) are given here for the record:
\begin{figure}[H]
    \begin{subfigure}[h]{0.38\textwidth}
    \centering
        \begin{tabular}{|cc|}
         \cline{1-2}
          BS$_1$: & \es{INV / r}                 \\       
          BS$_2$: & {\es{f.}\es{INV}}{ / \es{r}} \\
         BS$_3$: & \es{Dep (Current).INV / r}              \\
         \hline
         \hline
         BC$_1$: & \es{INV} \e{/ r}    \\
         BC$_2$: & {\es{a.}\es{INV}}{ / \es{r}} \\ 
         BC$_3$ &  \es{Dep (Current).INV / r} \\
         \cline{1-2}
        \end{tabular}
        \caption*{\Rule{Strong-pure}}
        \label{tab:strong_query}
    \end{subfigure}
\hfill
    \begin{subfigure}[h]{0.3\textwidth}
    \centering
        \begin{tabular}{|cc|}
             \hhline{-|-}
             BS$_1$: & \es{INV / r}                \\
             BS$_2$: & {\es{f.}\es{INV}}{ / \es{r}} \\

             \hline
             \hline
              BC$_1$: & \es{INV} \es{/ r}                         \\
              BC$_2$: & {\es{a.}\es{INV}}{ / \es{r}} \\

             BC$_3$: & \es{x.INV\_REM / r}                              \\
             \cline{1-2}
        \end{tabular}
        \caption*{\Rule{Weak-C-pure}}
        \label{tab:weak_c_query}
    \end{subfigure}
\hfill
    \begin{subfigure}[h]{0.25\textwidth}
    \centering
        \begin{tabular}{|cc|}
             \hhline{-|-}
             BS$_1$: & \es{INV_LOC / r}           \\
              BS$_2$: & {\es{f.}\es{INV}}{ / \es{r}} \\

             \hline
             \hline
             BC$_1$: &   \es{INV} \es{/ r}                             \\
             BC$_2$: & {\es{a.}\es{INV}}{ / \es{r}} \\
             \cline{1-2}
        \end{tabular}
        \caption*{\Rule{Weak-S-pure}}
        \label{tab:weak_s_query}
    \end{subfigure}
\end{figure}

\noindent Applying these rules assumes that the verifier can ascertain that a query is pure, either through a language rule or by static analysis. Both approaches have received considerable attention but fall beyond the scope of the present work.

\subsection{Concurrency} \label{initialization_queries_concurrency:concurrency}

This discussion has assumed a sequential programming context.
In an unbridled concurrent framework, for example using multithreading, there is no hope for a meaningful notion of class invariant, since furtive access is ever looming. The risk of an Ethereum-DAO-like failure is consequently high. 

Preserving the notion of class invariant, and the associated reasoning capabilities, requires a more controlled concurrent mechanism.
In SCOOP \cite{eiffel_software_concurrent_nodate}, all access to shared objects is exclusive; then the entire reasoning and rules of this article apply unchanged.

\subsection{Retained rules} \label{initialization_queries_concurrency:retained_rules}

For expository purposes, this article has introduced the rules as successive versions.
The final result, however, is precisely defined. The rules to be applied for verification are:
\begin{itemize}
    \item For commands (procedures), \Rule{Weak-S} (expressed in full in traditional Hoare-style inference-rule notation in the preview, section \ref{progression}).
    \item For creation procedures, \Rule{Weak-S-Creation}.
    \item For pure queries, choice (per the criteria discussed in \ref{initialization_queries_concurrency:pure_queries}) of programming style : \Rule{Weak-C-pure} or \Rule{Weak-S-pure}.
\end{itemize}

\section{Challenge problems and solutions} \label{challenge_problems_and_solutions}

The following discussion reports on the application of the rule to a  number of problems widely considered in the literature to embody the most tricky challenges to invariant-based verification. All raise invariant violations when handled with a simple notion of invariant.

Some have been verified before, but only with extra programmer annotations.
The versions used here are annotation-free, although some have required a modicum of tweaking of the code, as is often the case for verification. All source code for the examples is available (with better formatting than the compact forms given below to save space) in a GitHub repository \cite{supporting_material}.

As the ongoing integration of the proof rule into a proof tool (AutoProof \cite{furia_autoproof_2017}) is not complete, verification has consisted of one or both of:
\begin{itemize}
    \item Manual application of the rules.
    \item Dynamic testing, taking advantage of Eiffel's assertion constructs (including class invariants as well as routine pre- and postconditions), and the mechanisms in the associated development environment (EiffelStudio) for monitoring these assertions at run time, widely used for testing and debugging.
\end{itemize}
While not as final as mechanized static proofs of correctness, these approaches are meaningful since:
\begin{itemize}
    \item The examples, while tricky, are small, making manual proofs credible.
    \item The state space (for cases that could trigger violations) is typically small, so dynamic monitoring can cover all relevant scenarios.
\end{itemize}

\noindent These efforts should be viewed as stepping stones towards the goal of mechanized static proofs of correctness.

\subsection{Marriage, non-recursive} \label{marriage_non_recursive}

The class  text with contracts is the following (in the interest of space the formatting does not respect indentation rules, and the creation procedure is omitted):

\begin{lstlisting}[numbers=left, firstnumber=auto]
class PERSON feature spouse: PERSON; is_married: BOOLEAN
  marry (o: PERSON)             -- Marry current person to `o'.
    require pre$_1$: o $\neq$ Current; pre$_2$: $\lnot$ is_married; pre$_3$: $\lnot$ o.is_married
         do set_spouse (o)
            set_married
            o.set_spouse (Current)
            o.set_married
     ensure post$_1$: o.spouse = Current; post$_2$: spouse = o end
  divorce (o: PERSON)           -- Remove marriage link with `o'.
    require pre$_1$: is_married; pre$_2$: spouse.is_married; pre$_3$: o=spouse
         do o.unset_married
            unset_married
            o.unset_spouse
            unset_spouse
     ensure post$_1$: $\lnot$ is_married; post$_2$: $\lnot$ (old spouse).is_married end
feature {PERSON}
  set_married do is_married := True ensure post$_1$: is_married end
  unset_married do is_married := False ensure post$_1$: $\lnot$ is_married end
  set_spouse (o: PERSON)            -- Make `o' the spouse of current person.
    require pre$_1$: o $\neq$ Current; pre$_2$: $\lnot$ o.is_married $\Rightarrow$ (o.spouse = Void)
            pre$_3$: o.is_married $\Rightarrow$ (o.spouse = Current)
         do spouse := o
     ensure post$_1$: spouse = o end
  unset_spouse do spouse := Void ensure post$_1$: spouse = Void end
invariant inv$_1$: spouse $\neq$ Current; inv$_2$: is_married = (spouse $\neq$ Void)
          inv$_3$: is_married $\Rightarrow$ (spouse.spouse = Current) end
\end{lstlisting}

\noindent (Routine \e{divorce} has an argument --- which must equal to  \e{spouse} --- because of the Selective Export Call Rule (\ref{slicing_condition}), but it is easy to wrap it into a routine with no argument, as done with \e{insert_right} in the next example.) The proof appears as (a), (b) and (c) below. For conciseness, such proofs use a tabular notation, with one row per  goal to be proven.
A checkmark $\checkmark$ signals an available fact. A numeric prefix refers to a line number in the code: for the preceding code, \e{4.pre$_3$} denotes the precondition clause \e{pre$_3$} of the feature (\e{set_spouse}) called at line 4.
If there is no such prefix, the fact is a contract element or an AS/BS component from the proof rule; for example, \e{AS$_1$.o.inv$_1$} states that \e{inv$_1$} holds for \e{o}, satisfying entry AS$_1$ of the rule.
\begin{table}[H]
    \begin{subtable}[h]{0.55\textwidth}
    \centering
    \begin{tabular}{|l||c|c|c|c|c|c|c|}
        \hline
           \diagbox[innerwidth=2cm]{\textbf{\small{Goals}}}{\textbf{\small{Facts}}} &
           \rot{ \es{BS$_2$.o.inv$_2$} } &
           \rot{\es{pre$_1$}} &
           \rot{\es{pre$_3$}} &
           \rot{\es{4.post$_1$}} &
           \rot{\es{5.post$_1$}} &
           \rot{\es{6.post$_1$}} &
           \rot{\es{7.post$_1$}} \\
        \hline
        \hline
        \es{4.pre$_1$}, \es{6.pre$_1$} & & \checkmark & & & &  &  \\
        \hline
        \es{4.pre$_2$} & \checkmark & & \checkmark & & &  & \\
        \hline
        \es{4.pre$_3$} & & & \checkmark & & &  &  \\    
        \hline
        \es{6.pre$_2$} & & & & & \checkmark &  &  \\
        \hline
        \es{6.pre$_3$} & & & & \checkmark & &  &  \\
        \hline
        \es{post$_1$} & & & & & & \checkmark  &  \\
        \hline
        \es{post$_2$} & & & & \checkmark & &  &  \\
        \hline
        \es{AS$_1$.inv$_1$} & & \checkmark & & \checkmark & &  &  \\
        \hline
        \es{AS$_1$.inv$_2$} & & & & \checkmark & \checkmark &  &  \\
        \hline
        \es{AS$_1$.inv$_3$} & & & & \checkmark & \checkmark & \checkmark  &  \\
        \hline
        \es{AS$_2$.o.inv$_1$} & & \checkmark & & & & \checkmark  &  \\
        \hline
        \es{AS$_2$.o.inv$_2$} & & & & & & \checkmark &  \checkmark  \\
        \hline
        \es{AS$_2$.o.inv$_3$} & & & &  \checkmark & & \checkmark & \checkmark 
        \\
        \hline
    \end{tabular}
    \caption{Proof of \es{marry}.}
    \label{marry:proof}
    \end{subtable}
\hfill
\begin{tabular}{c}
    \begin{subtable}[t]{0.4\textwidth}
    \centering
    \begin{tabular}{|l||c|c|c|c|}
        \hline
           \diagbox[innerwidth=1.6cm]{\textbf{\small{Goals}}}{\textbf{\small{Facts}}} &
           \rot{\es{pre$_1$}} &
           \rot{\es{pre$_2$}} &
           \rot{\es{pre$_3$}} &
           \rot{\es{22}} \\
        \hline
        \hline
        \es{post$_1$} & & & & \checkmark \\
        \hline
        \es{AS$_2$.o.inv$_1$} & \checkmark & \checkmark & \checkmark & \\
        \hline
        \es{AS$_2$.o.inv$_2$} & & \checkmark & \checkmark & \\
        \hline
        \es{AS$_2$.o.inv$_3$} & & & \checkmark & \checkmark \\
        \hline
    \end{tabular}
    \caption{Proof of \es{set_spouse}.}
    \label{set_spouse:proof}
    \end{subtable}
\\
    \begin{subtable}[b]{0.4\textwidth}
    \centering
    \begin{tabular}{|l||c|c|c|c|}
        \hline
           \diagbox[innerwidth=2.5cm]{\textbf{\small{Goals}}}{\textbf{\small{Facts}}} &
           \rot{\es{11.post$_1$}} &
           \rot{\es{12.post$_1$}} &
           \rot{\es{13.post$_1$}} &
           \rot{ \es{14.post$_1$} } \\
        \hline
        \hline
        \es{post$_1$}, \es{AS$_1$.inv$_3$} & & \checkmark & & \\
        \hline
        \es{post$_2$}, \es{AS$_1$.o.inv$_3$} & \checkmark & & & \\
        \hline
        \es{AS$_1$.inv$_1$} & & & & \checkmark \\
        \hline
        \es{AS$_1$.inv$_2$} & & \checkmark & & \checkmark \\
        \hline
        \es{AS$_1$.o.inv$_1$} & & & \checkmark & \\
        \hline
        \es{AS$_1$.o.inv$_2$} & \checkmark & & \checkmark & \\
        \hline
    \end{tabular}
    \caption{Proof of \es{divorce}.}
    \label{divorce:proof}
    \end{subtable}
\end{tabular}
\end{table}

\subsection{Circular doubly-linked list} \label{circular_list}

The full program text appears below. The body of \e{make} only consists of assignments and trivially establishes both its postcondition and the class invariant.

Verification of \e{set_left} and \e{set_right} is straightforward too --- the assignments establish the postconditions, and the preconditions establish the invariants on exit. 
The generic parameter \e{G} represents the type of list elements.

\noindent Note that routine \e{insert_between_two} is ``wrapped'' by a public routine \e{insert_right} as the extra arguments of the first should be in internally called (secret) routine, so that the preconditions will be guaranteed.

\begin{lstlisting}[numbers=left, firstnumber=auto]
class CIRCULAR_NODE [G] create make feature {CIRCULAR_NODE}
  make (v: G)            -- Initialize with single node of value `v'.
        do value := v
           left := Current
           right := Current
    ensure post$_1$: value = v; post$_2$: left = Current; post$_3$: right = Current end
  set_right (o: CIRCULAR_NODE [G])      -- Add `o' as right neighbor.
    require pre$_1$: o.left.right = o; pre$_2$: o.right.left = o
         do right := o ensure post$_1$: right = o end
  set_left (o: CIRCULAR_NODE [G])      -- Add `o' as left neighbor.
    require pre$_1$: o.left.right = o; pre$_2$: o.right.left = o
         do left := o ensure post$_1$: left = o end
  insert_between_two (v: G; l, r: CIRCULAR_NODE[G])  -- Add node of value `v' between 'l' and 'r'.
    require pre$_1$: l.right = r; pre$_2$: r.left = l;
         do make (v)
            l.set_right (Current)
            r.set_left (Current)
            left := l
            right := r
     ensure post$_1$: value = v; post$_2$: left = l; post$_3$: right = r end
feature value: G; left, right: CIRCULAR_NODE[G]
  remove (l: CIRCULAR_NODE[G])      -- Delete `l' from its list.
  	require pre$_1$: l = left;
            pre$_2$: l.left.left.right = l.left; pre$_3$: l.left.right.left = l.left;
            pre$_4$: right.left.right = right; pre$_5$: right.right.left = right; 
         do if l = right then l.make (value)
            else l.insert_between_two (l.value, l.left, right) end
            make (value)
     ensure post$_1$: left = Current; post$_2$: (old left).right = old right end
  insert_right (v: G; l: CIRCULAR_NODE[G]) -- Insert node of value 'v' on the right of 'l'
     do insert_between_two (v, l, l.right) end
invariant inv$_1$: left.right = Current; inv$_2$: right.left = Current end
\end{lstlisting}

The proofs of \e{insert}  and \e{remove} appear below.

\begin{table}[H]
    \centering
    \begin{tabular}{|l||l|l|l|l|l|l|l|l|l|}
         \cline{1-10}
           \diagbox[innerwidth=5cm]{\textbf{\small{Goals}}}{\textbf{\small{Facts}}} &
           \rot{\es{BS$_2$.l.inv$_1$}} &
           \rot{ \es{BS$_2$.r.inv$_2$} } &
           \rot{\es{15.post$_1$}} &
           \rot{\es{15.post$_2$}} &
           \rot{\es{15.post$_3$}} &
           \rot{\es{16.post$_1$}} &
           \rot{\es{17.post$_1$}} &
           \rot{\es{18}} &
           \rot{\es{19}} \\
        \cline{1-10}
        \cline{1-10}
        \es{16.pre$_1$}, \es{16.pre$_2$}, \es{17.pre$_1$}, \es{17.pre$_2$}  & & & & \checkmark & \checkmark & & & & \\
        \cline{1-10}
        \es{post$_1$} & & & \checkmark & & & & & & \\
        \cline{1-10}
        \es{post$_2$} & & & & & & & & \checkmark & \\
        \cline{1-10}
        \es{post$_3$} & & & & & & & & & \checkmark  \\
        \cline{1-10}
        \es{AS$_2$.l.inv$_1$} & \checkmark & & & & & & & & \\
        \cline{1-10}
        \es{AS$_2$.l.inv$_2$}, \es{AS$_1$.inv$_1$}  & & & & & & \checkmark & & \checkmark & \\
        \cline{1-10}
        \es{AS$_2$.r.inv$_1$}, \es{AS$_1$.inv$_2$} & & & & & & & \checkmark & & \checkmark \\
        \cline{1-10}
        \es{AS$_2$.r.inv$_2$} & & \checkmark & & & & & & & \\
        \cline{1-10}
    \end{tabular}
    \caption*{Proof of \es{insert_between_two}}
\end{table}
\begin{table}[H]
    \centering
    \begin{tabular}{|l||c|c|c|c|c|c|c|c|c|c|c|}
        \cline{1-12}
           \diagbox[innerwidth=3cm]{\textbf{\small{Goals}}}{\textbf{\small{Facts}}} &
           \rot{\es{pre$_1$}} &
           \rot{\es{pre$_2$}} &
           \rot{\es{pre$_3$}} &
           \rot{\es{pre$_4$}} &
           \rot{\es{pre$_5$}} &
           \rot{\es{26.post$_3$}} &
           \rot{\es{27.post$_3$}} &
           \rot{\es{27.AC$_1$.inv$_1$}} &
           \rot{ \es{27.AC$_1$.inv$_2$} } &
           \rot{\es{28.post$_2$}} &
           \rot{\es{28.post$_3$}} \\
        \cline{1-12}
        \cline{1-12}
        \es{27.BC$_2$.l.left.inv$_1$} & & \checkmark & & & & & & & & & \\
        \cline{1-12}
        \es{27.BC$_2$.l.left.inv$_2$} & & & \checkmark & & & & & & & & \\
        \cline{1-12}
        \es{27.BC$_2$.right.inv$_1$} & & & & \checkmark & & & & & & & \\
        \cline{1-12}
        \es{27.BC$_2$.right.inv$_2$} & & & & & \checkmark & & & & & & \\
        \cline{1-12}
        \es{post$_1$} & & & & & & & & & & \checkmark & \\
        \cline{1-12}
        \es{post$_2$} & \checkmark & & & & & \checkmark & \checkmark & & & & \\
        \cline{1-12}
        \es{AS$_1$.inv$_1$}, \es{AS$_1$.inv$_2$} & & & & & & & & & & \checkmark & \checkmark \\
        \cline{1-12}
        \es{AS$_2$.l.inv$_1$} & & & & & & & & \checkmark & & \checkmark & \checkmark \\
        \cline{1-12}
        \es{AS$_2$.l.inv$_2$} & & & & & & & & & \checkmark & \checkmark & \checkmark \\      
        \cline{1-12}
    \end{tabular}
    \caption*{Proof of \es{remove}}
\end{table}

\subsection{Observer} \label{challenge_problems_and_solutions:observer}


Class \e{OBSERVER} is a simple implementation of the Observer pattern.

\begin{lstlisting}[numbers=left, firstnumber=auto]
class OBSERVER feature {NONE}  -- Features exported to no class
  make (s: SUBJECT)
         do subject := s;
            s.register_observer(Current)
     ensure post$_1$: subject = s; post$_2$: s.subscribers.has(Current)
            post$_3$: old s.subscribers.count + 1 = s.subscribers.count end
feature {SUBJECT, OBSERVER}
  update
        do cache := subject.value
    ensure post$_1$: cache = subject.value end
feature cache: INTEGER; subject: SUBJECT
  set_subject (s: SUBJECT)
        do subject.subscribers.search(Current)
           subject.subscribers.remove
           subject := s
           s.register_observer(Current)
    ensure post$_1$: subject = s end
invariant inv$_1$: cache > 0; inv$_2$: cache = subject.value;
          inv$_3$: subject.subscribers.has (Current) end
\end{lstlisting}

\noindent The proof of \e{OBSERVER.update} is immediate: the postcondition follows from the assignment, and the \e{AS$_1$.Current.inv$_3$} obligation from assuming \e{BS$_1$.Current.inv$_3$}.

\begin{table}[H]
    \begin{subtable}[h]{0.4\textwidth}
    \centering
    \begin{tabular}{|l||c|c|c|c|c|c|c|c|c|}
        \hline
           \diagbox[innerwidth=1.5cm]{\textbf{\small{Goals}}}{\textbf{\small{Facts}}} &
           \rot{\es{15}} &
           \rot{\es{BS$_2$.s.inv$_1$}} &
           \rot{ \es{16.AC$_2$.inv$_2$} } &
           \rot{\es{16.post$_1$}} &
           \rot{\es{BS$_2$.s.inv$_3$}} &
           \rot{\es{BS$_2$.s.inv$_2$}} &
           \rot{\es{16.AC$_2$.INV}} \\
        \hline
        \hline
        \es{16.pre$_1$} & \checkmark & & & & &  & \\
        \hline
        \es{16.pre$_2$} & \checkmark & \checkmark & & & &  & \\
        \hline
        \es{post$_1$} & \checkmark & & & & &  & \\
        \hline
        \es{AS$_1$.INV} & & & & & & &  \checkmark  \\
        \hline
        \es{AS$_2$.s.inv$_1$} & & \checkmark & & & &  & \\
        \hline
        \es{AS$_2$.s.inv$_2$} & \checkmark & & \checkmark & \checkmark & & \checkmark  & \\
        \hline
        \es{AS$_2$.s.inv$_3$} & \checkmark & & & \checkmark & \checkmark &  & \\
        \hline
    \end{tabular}
    \caption*{Proof of \es{set_subject}.}
    \label{set_subject:proof}
    \end{subtable}
\hfill
    \begin{subtable}[h]{0.5\textwidth}
    \centering
    \begin{tabular}{|l||c|c|c|c|c|c|c|c|}
        \hline
           \diagbox[innerwidth=1.9cm]{\textbf{\small{Goals}}}{\textbf{\small{Facts}}} &
           \rot{\es{3}} &
           \rot{\es{BS$_2$.s.inv$_1$}} &
           \rot{\es{4.post$_1$}} &
           \rot{\es{4.post$_2$}} &
           \rot{ \es{4.AC.$_2$.inv$_2$} } &
           \rot{\es{BS$_2$.s.inv$_2$}} &
           \rot{\es{BS$_2$.s.inv$_3$}} &
           \rot{\es{4.AC$_2$.INV}} \\
        \hline
        \hline
        \es{4.pre$_1$}, \es{post$_1$} & \checkmark & & & & & & & \\
        \hline
        \es{4.pre$_2$} & \checkmark & \checkmark & & & & & & \\
        \hline
        \es{post$_2$} & & & \checkmark & & & & & \\
        \hline
        \es{post$_3$} & & & & \checkmark & & & & \\
        \hline
        \es{AS$_1$.INV} & & & & & & & & \checkmark \\
        \hline
        \es{AS$_2$.s.inv$_1$} & & \checkmark & & & & & &  \\
        \hline
        \es{AS$_2$.s.inv$_2$} & \checkmark & & \checkmark & & \checkmark & \checkmark & & \\
        \hline
        \es{AS$_2$.s.inv$_3$} & \checkmark & & \checkmark & & & & \checkmark & \\
        \hline
    \end{tabular}
    \caption*{Proof of \es{OBSERVER.make}.}
    \label{observer_make:proof}
    \end{subtable}
\end{table}

\subsection{Subject} \label{challenge_problems_and_solutions:subject}

The code for the \e{Subject} in the Observer pattern is as follows.

\begin{lstlisting}[numbers=left, firstnumber=auto]
class SUBJECT create make feature {NONE}
  make (v: INTEGER)             -- Initialize with observed value `v'.
    require pre$_1$: v > 0
         do value := v
            create {ARRAYED_LIST [OBSERVER]} subscribers.make(0)
     ensure post$_1$: value = v end
  update_observer (o: OBSERVER)     -- Set `o' as the observer.
    require pre$_1$: subscribers.has(o); pre$_2$: o.subject = Current; pre$_3$: value > 0
         do o.update
     ensure post$_1$: o.cache = value end
feature {OBSERVER}
  register_observer (o: OBSERVER)   -- Allow `o' to become the observer.
    require pre$_1$: o.subject = Current; pre$_2$: value > 0
         do subscribers.extend(o)
            o.update 
     ensure post$_1$: subscribers.has(o)
            post$_2$: old subscribers.count + 1 = subscribers.count end
feature value: INTEGER; subscribers: LIST [OBSERVER]
  update_value (v: INTEGER)     -- Set the observed value to `v'.
    require pre$_1$: v > 0; pre$_2$: $\forall$o: subscribers | o.subject = Current
         do value := v
            across subscribers as o loop update_observer(o) end
     ensure post$_1$: value = v end
invariant   inv$_1$: value > 0; inv$_2$: $\forall$o: subscribers | o.cache = value;
            inv$_3$: $\forall$o: subscribers | o.subject = Current end
\end{lstlisting}

\noindent Proving \e{make} and \e{update_observer} is trivial: all the proof goals imposed by the proof rule directly follow from these features' preconditions and implementations, so we do not present them.

\begin{table}[h]
    \begin{subtable}[h]{0.4\textwidth}
    \centering
    \begin{tabular}{|l||c|c|c|c|}
        \hline
           \diagbox[innerwidth=2cm]{\textbf{\small{Goals}}}{\textbf{\small{Facts}}} &
           \rot{\es{pre$_1$}} &
           \rot{\es{pre$_2$}} &
           \rot{\es{21}} &
           \rot{ \es{22.post$_1$} } \\
        \hline
        \hline
        \es{post$_1$} & & & \checkmark & \\
        \hline
        \es{AS$_1$.inv$_1$} & \checkmark & & \checkmark &  \\
        \hline
        \es{AS$_1$.inv$_2$} & & \checkmark & & \checkmark \\
        \hline
        \es{AS$_1$.inv$_3$} & & \checkmark & & \\
        \hline
    \end{tabular}
    \caption*{Proof of \es{update_value}.}
    \label{update_value:proof}
    \end{subtable}
\hfill
    \begin{subtable}[h]{0.55\textwidth}
    \centering
    \begin{tabular}{|l||c|c|c|c|c|}
        \hline
           \diagbox[innerwidth=2.0cm]{\textbf{\small{Goals}}}{\textbf{\small{Facts}}} &
           \rot{\es{pre$_1$}} &
           \rot{\es{pre$_2$}} &
           \rot{ \es{14.item_inserted} } &
           \rot{\es{14.new_count}} &
           \rot{\es{15.post$_1$}} \\
        \hline
        \hline
        \es{post$_1$} & & & \checkmark & & \\
        \hline
        \es{post$_2$} & & & & \checkmark & \\
        \hline
        \es{AS$_2$.o.inv$_1$} & \checkmark & \checkmark & & & \checkmark \\
        \hline
        \es{AS$_2$.o.inv$_2$} & & & & & \checkmark \\
        \hline
        \es{AS$_2$.o.inv$_3$} & \checkmark & & \checkmark & & \\
        \hline
    \end{tabular}
    \caption*{Proof of \es{register_observer}.}
    \label{register_observer:proof}
    \end{subtable}
\end{table}

\section{Previous work} \label{related_work}
The first presentation of class invariants is a single mention of ``\emph{invariant of the class}'' in Hoare's 1972 ``correctness of data representations'' \cite{hoare_proof_1972}.
Experimental 1970s languages included representation invariants: Alphard \cite{wulf_introduction_1976} provides both abstract invariants (defined on the underlying Abstract Data Type or ADT) and representation invariants (on a specific implementation).
OO design provides the opportunity to use as many levels as needed instead of just two. Abstraction and representation become relative concepts; each class inherits its parents' invariants, adding its own. The development of the concept of class invariant for OO appeared in 1985 in \cite{meyer1985,meyer_eiffel_1988} and other Eiffel-related publications.
\cite{oosc_1,oosc_2} (1988, 1997) furthered the concepts.
Other formalisms with invariant support include JML \cite{burdy_overview_2005} and Spec\# \cite{barnett_spec_2005}.

The problem of precise semantics for invariants has attracted considerable attention. \cite{nelson_verifying_1983} introduced a reachability predicate for linear lists. \cite{skevoulis_generic_2000} developed an invariant-based tool to analyze Java code for such  defects as illegal dereferencing. \cite{malloy_exploiting_2006} exploited design patterns to validate OCL \cite{warmer_object_2003}  invariants in C++.

 The issue of reference leak, as analyzed in the present work, arises when  operations on an object may invalidate the invariant of another object. One category of solutions has received particular attention: \textit{ownership} techniques, in which some objects are ``owned'' by others, with the rule that any operation on an owned object must go through its owner, so as to control invariant change. Programmers are responsible for declaring ownership relations in the code. This idea has many variants: 

\begin{itemize}
    \item The methodology of \cite{huizing_verification_2000} uses the notion of ``vulnerable'' objects, meaning objects that may be invalidated by operations on instances of another class. The code must declare \emph{logical variables} to identify vulnerable objects.

    \item   The Universe Type System \cite{muller_modular_2002} distributes objects into ownership contexts limiting updates. \cite{barnett_verification_2004} introduced an ownership-based mechanism that includes boolean states in which the invariant is known to hold.
    
    \item
   Later work \color{black} went further through a \emph{friendship} protocol to allow state dependence across ownership boundaries \cite{barnett_friends_2004}, a  case of \emph{cooperation-based} approaches.  \cite{middelkoop_cooperation-based_2006,middelkoop_invariants_2008}, also known as \emph{visibility-based} \cite{muller_modular_2006}, formalized and proved sound in \cite{naumann_towards_2006}. 
   
    \item The Boogie methodology \cite{leino_object_2004}  refined the ideas by removing bounds on the number of objects in an ownership context and letting objects change their contexts. \cite{jacobs_safe_2005} extended the methodology to the multithreaded case by protecting object structures from race conditions. Another extension \cite{leino_modular_2005} handles invariants over static fields. An extension \cite{leino_class-local_2008} adds yet another modifier, \emph{additive}, specifying which fields may be mentioned in subclass invariants.

    \item In an ownership context, \cite{naumann_assertion-based_2005} used \emph{ghost} fields to handle reentrant method calls and shared references.
\cite{muller_modular_2006} combined ownership and visibility techniques for modular specification and verification of layered structures. New extensions to the Boogie approach use \emph{history invariants} --- two-state invariants describing the evolution of data values --- to verify variations of the Observer pattern; one of the latest, 
\emph{semantic collaboration} \cite{polikarpova_flexible_2014}, enhances the methodology with such ghost fields as \emph{subjects} and \emph{observers}. 

    \item \cite{lu_validity_2007} introduces two sets of objects: those in the \emph{validity invariant} must be valid before and after, unlike those in the \emph{validity effect}. On exit, re-validation is only needed for the intersection. Later work \cite{huster_more_2014} generates the sets automatically from visibility modifiers.
    
    \item \cite{balzer_modular_2010} addresses invariant issues in a language with contracts, treating inter-object relations as first-class citizens.
Verification relies on a ``Matreshka Principle'' to guarantee absence of transitive call-backs and restore a visible-states semantics for multi-object invariants \cite{balzer_verifying_2011}. The methodology involves annotations to declare relationships and their participants.

    \item \cite{banerjee_regional_2008,banerjee_local_2013,banerjee_local_2013-1} introduce region logic supporting heap-local reasoning about mutation and separation via ghost fields and variables of type ``region'' consisting of finite sets of object references. 
\cite{bao_jml_2018} adapted region logic to JML and extended it with separating conjunction \cite{reynolds_separation_2002}.

    \item \cite{cohen_local_2010} proposed two-state \emph{locally checked invariants} (LCI), reducing verification of a concurrent program to checking that they hold for every  update.

\end{itemize}

\noindent In addition to handling reference leak, the approaches cited generally address furtive access through  special instructions \emph{wrap} and \emph{unwrap} (sometimes \emph{pack} and \emph{unpack}) to specify when objects must satisfy their invariants and when they are permitted to violate them, for example inside the execution of a routine. Such instructions have no effect on program execution; in other words, they are intended for the verifier, not the compiler.

\cite{drossopoulou_unified_2008} proposes 7 verification parameters and conditions on them to guarantee soundness; analyzing existing approaches in terms of the proposed semantics. \cite{meyer_class_2016} was an initial iteration towards the present work.

Unlike the static approaches reviewed so far, other work performs \textit{dynamic} verification of programs, requiring execution:

\begin{itemize}
    \item \cite{gopinathan_runtime_2008} proposed a run-time verification scheme guaranteeing that any invariant violation is detected exactly where it occurs, and proved its correctness. The scheme automatically tracks dependencies between objects. 
    
    \item  \cite{gorbovitski_efficient_2008} presented another framework for run-time invariant checking.
    
    \item \cite{christakis_synthesizing_2014}, an extension to Pex \cite{hutchison_pexwhite_2008}, synthesizes parameterized unit tests \cite{tillmann_parameterized_2005} from multi-object invariants.
    
\end{itemize}

\noindent The Ethereum DAO bug \cite{dao_bug}, due to a callback \cite{Sirer_Ethereum_2016}, gave rise to invariant-based methodologies for smart contracts \cite{grossman_online_2018,aiello_call_me_2020,albert_taming_2020}. \cite{aiello_call_me_2020} adds a type invariant to smart contracts. \cite{grossman_online_2018} introduces the notion of Effectively Callback Free (ECF) objects (as in section \ref{smart} of the present work). \cite{albert_taming_2020} extended the concept and provided a static analysis technique to ensure effective-callback-freedom.

\section{Limitations, future work and conclusion} \label{conclusion}

The preceding discussion does not consider inheritance. Together with associated techniques of polymorphism and dynamic binding, inheritance is an integral part of object-oriented programming. Under inheritance, invariants accumulate \cite{meyer1985,oosc_1}. We surmise that the rules given will naturally extend to inheritance, but the present study has not yet explored this matter in depth. Also, not considered in the current state of this work include reflection and mechanisms for passing routines as arguments to other routines, as either plain function pointers (in C++) or  more abstract values such as Eiffel's agents and C\#'s delegates. Such arguments can cause callbacks, which the analysis must detect.

The ``weak'' rules currently retained for addressing reference leak require explicit enforcement of the ``remote'' part of the invariant, preferably on the supplier side, and may require some fine-tuning of existing code for verification. The ``strong'' version of the rules would avoid these problems, but making them convenient in practice raises the question of how to compute the dependent sets statically, efficiently, and modularly -- a question that remains open. 

Proofs of properties of the underlying theory (consistency of successive versions of Global Consistency, and associated theorems) have been conducted manually as reported above, in the usual style of mathematical proofs, and have been checked carefully, but have not been subject to an automated theorem prover. As to the application of these rules to programs, the process of integration into an automatic prover, AutoProof, is not complete.

While the rules do not assume any specific typing system, or even that the language is statically typed, they do assume that it is reasonably well-behaved, for example that it prohibits a routine from modifying its own arguments (\ref{slicing_condition}). It may be possible to generalize the approach to arbitrary, undisciplined languages, but we have not attempted such a feat.

The Selective Export Call Rule (\ref{selective_exports}), requiring the targets of calls to selectively exported routines to be formal arguments, may require some wrapping effort, as in the verification of \e{divorce} in \ref{marriage_non_recursive} and \e{insert_right} in \ref{circular_list}. As discussed in section \ref{smart}, the burden appears acceptable in practice, and disappears if the analysis can detect the absence of callbacks. 

As mentioned in the discussion of the Callback Model (end of section \ref{callbacks}), it would be useful to conduct an empirical study of a large OO code base, using a prover such as AutoProof, to find out whether most qualified calls to non-library routines do already wrap up the client object (make it consistent) first, meeting the \e{INV} or \e{INV / r} part of the call's precondition.

With these qualifications, the rules as they stand provide an answer to the goals set at the beginning of this article: opening up an effective way to use class invariants, based on an actual axiomatization, not just a ``methodology''; identifying the precise obstacles (three distinct threats) to the soundness of the basic invariant concept; and providing solutions to all of them, without requiring of the programmer any annotation burden.    

The rules give practicing software developers a basis for making the fundamental notion of class invariant a key tool of their work, helping them benefit from the full power of OO concepts. Our own experience as programmers shows that extensive use of invariants leads to better programs, easier to write, understand and get right (and debug when not). We are, as noted, integrating the rules of this article into an automatic prover, part of a determined effort to restore the belief in a sanity clause.

\noindent\rule{12.2cm}{0.4pt}

\noindent \textit{Acknowledgments} We are grateful to the referees for many detailed and perceptive comments which led to considerable improvements of the article in successive revisions. Iulian Ober, now with ISAE-Supaéro in Toulouse, made important suggestions on an early version of the work. A presentation of an intermediate stage of the research, at a meeting of the IFIP WG2.3 working group in Villebrumier, elicited fruitful insights from Gary Leavens and other members of the group, many of whom have themselves made important contributions to the topic over the years. The work benefited  from the ``semantic collaboration'' concepts \cite{polikarpova_flexible_2014} integrated into the original AutoProof, and discussions with its coauthors:  Carlo Furia, Nadia Polikarpova (who in an informal discussion brought up the observation that selective exports could help) and Julian Tschannen. Li Huang from Constructor Institute participated in a number of discussions. Alexandr Naumchev, formerly from Constructor Institute, was involved in the first version.

\bibliographystyle{ACM-Reference-Format}
\bibliography{class-invariants}


\end{document}